\documentclass[10pt,journal,compsoc]{IEEEtran}
  
\usepackage{cite}
\usepackage{amsmath}
\usepackage[obeyspaces]{url}
\PassOptionsToPackage{obeyspaces}{url}
\usepackage{hyperref}

 \newcommand{\new}[1]{#1}

\usepackage{xspace}
\def\acronym{\textsc{InterEvo-TR}\xspace}

\usepackage{soul}   %% texto resaltado
\usepackage{tcolorbox}
\usepackage{amsthm}
\usepackage{multicol,multirow}

\theoremstyle{definition}

\usepackage{xcolor}
\definecolor{verde}{rgb}{0.25,0.5,0.35}
\definecolor{jpurple}{rgb}{0.5,0,0.35}
\definecolor{darkgreen}{rgb}{0.0, 0.2, 0.13}
\definecolor{codegreen}{rgb}{0,0.6,0}
\definecolor{codepurple}{rgb}{0.58,0,0.82}
\usepackage{listings}

\newcommand{\estiloJava}{
\lstset{
    language=Java,
    basicstyle=\fontsize{7}{8}\selectfont\ttfamily,
    keywordstyle=\color{blue}\bfseries,
    stringstyle=\color{codepurple},
    commentstyle=\color{codegreen},
    morecomment=[s][\color{blue}]{/**}{*/},
    extendedchars=true,
    showspaces=false,
    showstringspaces=false,
    numbers=none,
    numberstyle=\tiny,
    breaklines=true,
    breakautoindent=true,
    captionpos=b,
    xleftmargin=0pt,
    tabsize=2,
    frame=lines
}}

\usepackage{enumitem}

\begin{document}
\title{\acronym: Interactive Evolutionary Test Generation with Readability Assessment}

\author{Pedro~Delgado-P\'erez, %~\IEEEmembership{Member,~IEEE,}
        Aurora~Ram\'irez, %~\IEEEmembership{Fellow,~OSA,}
        Kevin~J.~Valle-G\'omez, 
        Inmaculada~Medina-Bulo, 
        and \\Jos\'e Ra\'ul~Romero%~\IEEEmembership{Life~Fellow,~IEEE}% <-this % stops a space
\IEEEcompsocitemizethanks{\IEEEcompsocthanksitem P. Delgado-P\'erez, K.J. Valle-G\'omez and I. Medina-Bulo are with University of C\'adiz, Spain. \protect\\
E-mail: \{pedro.delgado, kevin.valle, inmaculada.medina\}@uca.es
\IEEEcompsocthanksitem A. Ram\'irez and J.R. Romero are with University of C\'ordoba, Spain.\protect\\
E-mail: \{aramirez, jrromero\}@uco.es}%<-this % stops an unwanted space
\thanks{Manuscript published December 09, 2022 \\DOI: \url{https://doi.org/10.1109/TSE.2022.3227418}}}


\markboth{IEEE Transactions on Software Engineering}%
{P. Delgado-P\'erez \MakeLowercase{\textit{et al.}}: \acronym: Interactive Evolutionary Generation of Test Suites with Readability Assessment}



\IEEEtitleabstractindextext{%
\begin{abstract}
Automated test case generation has proven to be useful to reduce the usually high expenses of software testing. However, several studies have also noted the skepticism of testers regarding the comprehension of generated test suites when compared to manually designed ones. This fact suggests that involving testers in the test generation process could be helpful to increase their acceptance of automatically-produced test suites. In this paper, we propose incorporating interactive readability assessments made by a tester into EvoSuite, a widely-known evolutionary test generation tool. Our approach, \acronym, interacts with the tester at different moments during the search and shows different test cases covering the same coverage target for their subjective evaluation. The design of such an interactive approach involves a schedule of interaction, a method to diversify the selected targets, a plan to save and handle the readability values, and some mechanisms to customize the level of engagement in the revision, among other aspects. To analyze the potential and practicability of our proposal, we conduct a controlled experiment in which \new{39} participants, including academics, professional developers, and student collaborators, interact with \acronym. Our results show that the strategy to select and present intermediate results is effective for the purpose of readability assessment. Furthermore, the participants' actions and responses to a questionnaire allowed us to analyze the aspects influencing test code readability and the benefits and limitations of an interactive approach in the context of test case generation, paving the way for future developments based on interactivity.
\end{abstract}

}

\maketitle

\IEEEdisplaynontitleabstractindextext

% \fi
\IEEEpeerreviewmaketitle

%-----------------
% 1 - INTRODUCTION
%-----------------
\IEEEraisesectionheading{\section{Introduction}\label{sec:introduction}}

\IEEEPARstart{T}{esting} is a crucial activity in the lifecycle of any software project to increase its quality and ensure its maintainability. However, testing is a costly phase that requires a considerable amount of resources~\cite{Myers:2011:AST:2161638}. As the complexity of industrial systems grows, the design of effective test suites becomes a harder task and exhaustive manual testing is no longer possible. Current research in automated testing aims to overcome these limitations by developing efficient techniques~\cite{Anand2013}, among which search-based algorithms have been extensively studied in the context of test generation~\cite{Ali2010}.

In object-oriented systems, test case generation involves the design of test scenarios comprised of a sequence of method calls and the evaluation of the resulting execution (i.e., outputs and/or final object states)~\cite{Wappler2006}. The generation of effective test suites often requires a deep understanding of the classes under test, since it implies choosing inputs able to reveal possible faults, setting initial states for the objects, defining the right order of invocations and adding proper assertions. Search-based test case generation automatizes this process, usually guided by coverage criteria. EvoSuite~\cite{fraser2012mutation}, an open-source test generation tool, has become a reference in this area. In the unit testing competition at SBST 2021~\cite{Panichella2021}, EvoSuite obtained the highest overall score among the five competing tools, a number of tools that in turn reveals the maturity achieved in this field. Indeed, many authors have built their proposals on top of EvoSuite~\cite{Evers20} or inspired by this tool (e.g., \textsc{EvoMaster} for REST API testing~\cite{Arcuri2021}), thus adapting the search to different testing scenarios and seeking to improve the process. 

Despite the great advances, some studies reveal that current automated tools still present some limitations that hamper their practicability~\cite{arcuri2018experience,Rojas2017}. More specifically, existing tools still have difficulties detecting complex, real-world faults~\cite{almasi2017industrial} and testers are often reluctant to adopt the generated tests due to their lack of readability when compared to manually-designed ones~\cite{Shamshiri2018}. As a motivating example of the latter issue, consider the test cases internally generated by EvoSuite for class \texttt{ArrayIntList}\footnote{This class is part of the \emph{Commons Collections Primitives} and implements an ordered collection of int values backed by an array.} in Figure~\ref{fig:testcases}. All these three test cases cover the same test target: a line in the method \texttt{removeAtElement}. Both \texttt{test0} and \texttt{test1} have the same length (excluding assertions, generated in a post-processing step). However, it might be difficult for a tester to understand at a first glance that \texttt{test0} is actually traversing \texttt{removeAtElement} because \texttt{clear} indirectly invokes that method. Also, \texttt{test2} is a bit longer than the other two test cases, but the argument values or the sequence of actions might be more meaningful than those in \texttt{test1} (e.g., they use different versions of the method \texttt{add}). In this situation, some questions arise: what is the covered functionality by each test case? Which of the features in these tests are desirable for a tester? Which test looks more ``human-written''? In summary, which test case is more readable? It seems that there is not an easy answer to these questions. It is not straightforward to map an automatically-generated test suite with the code covered by each test case, and no test readability standard applies to every tester equally due to its subjective nature~\cite{Rojas2017}. Contrarily, test readability is more about the understanding and preferences of testers, or even the guidelines followed in each organization. 

\begin{figure}[ht]
\begin{small}
\estiloJava
\begin{lstlisting}
@Test
public void test0()  throws Throwable  {
  ArrayIntList arrayIntList0 = new ArrayIntList();
  arrayIntList0.add(0, 0);
  arrayIntList0.clear();
  assertEquals(0, arrayIntList0.size());
}

@Test
public void test1()  throws Throwable  {
  ArrayIntList arrayIntList0 = new ArrayIntList(301);
  arrayIntList0.add(2181);
  int int0 = arrayIntList0.removeElementAt(0);
  assertEquals(0, arrayIntList0.size());
  assertEquals(2181, int0);
}

@Test
public void test2()  throws Throwable  {
  ArrayIntList arrayIntList0 = new ArrayIntList();
  arrayIntList0.add(0, 0);
  arrayIntList0.add(1, 1);
  int int0 = arrayIntList0.removeElementAt(1);
  assertEquals(1, arrayIntList0.size());
  assertEquals(1, int0);
}
\end{lstlisting}
\end{small}
\caption{Test cases generated with EvoSuite covering a line in the method \texttt{removeElementAt} in class \texttt{ArrayIntList}.}
\label{fig:testcases}
\end{figure}
 
Different automated techniques have recently contributed to dealing with the problem of the skepticism of testers regarding the comprehension of generated test suites by improving their final appearance~\cite{Daka2017,Panichella2016, Roy2020}. To address this challenge, however, the possibility of complementing current automated tools with the know-how of testers can represent a further significant step towards the acceptance of the generated tests. Indeed, incorporating human knowledge into the test generation process via interactive optimization has been recently mentioned as an appealing solution~\cite{Cohen2019, Rojas2017}. Interactive approaches allow the active participation of humans and have been explored for different problems in search-based software engineering (SBSE)~\cite{Ramirez2019,Perez2022, Rebai2022}. Focusing on search-based software testing (SBST), Marculescu et al. were the first to propose an interactive SBST tool for test data generation~\cite{Marculescu2015isbst}, which was successfully evaluated with an industrial partner~\cite{MarculescuFTP18}. Interactive optimization is especially well-suited when objectives are difficult to define~\cite{Marculescu:SBSE12}, or some additional subjective criteria need to be considered~\cite{Cohen2019}. Trying to improve the readability of test suites falls into this scenario~\cite{Rojas2017} since it is an abstract concept subject to the tester's own perception. 

To bridge the gap between testers and automated tools, this paper studies how interactive optimization could help address the challenge of generating more readable and meaningful test suites according to the tester's preferences. \new{In this approach, the concept of readability is intrinsically linked to the tester's choices, past experience and knowledge of the system. As such,
test readability is quantifiable from the tester's subjective perspective instead of from objective measures related to 
particular test code features (e.g., test length).} To this purpose, we adapt and extend EvoSuite to allow testers to incorporate their readability assessment \new{in the form of a readability score} and, thus, participate in the decision of which of the test cases generated are preferable to keep in relation to several coverage targets pursued in the search process. This includes, among other aspects, a schedule of interaction, a method to diversify the selected targets, a plan to save the readability value assigned to the inspected test cases and some mechanisms to customize the level of engagement in the revision of tests.

We applied this novel approach, called \acronym (Interactive Evolutionary Test generation with Readability assessment), in a study with \new{39} participants who interacted with the tool during a test generation process. The analysis of their actions and feedback offers interesting findings regarding the subjective perception of aspects affecting readability and the positive influence of design decisions in the interaction process (e.g., the target selection strategy or the number of programmed interventions). In terms of usefulness, our results suggest that \acronym not only helps reach more readable test suites based on the tester's preferences, but it also provides a space to reflect on the goals pursued by some of the test cases in the final test suite.

The unique contributions of our paper are:

\begin{itemize}
    \item An interactive approach specifically designed to integrate readability assessments into search-based test case generation.
    \item \acronym, the implementation of our interactive approach built on EvoSuite and available online\footnote{Replication package: \url{https://doi.org/10.5281/zenodo.7182195}}. This new version includes novel options to handle interactivity and readability assessment, which allows tailoring \emph{when} and \emph{how} interactions take place.
    \item An empirical study on the relation between test case length and the moment when test targets are covered, which serves to determine a strategy for the selection of more interesting targets from the perspective of readability assessment.
    \item An experimental study with humans to validate the feasibility of the interactive tool, which evaluates how participants addressed different aspects related to the approach: \emph{test readability}, \emph{interaction process} and \emph{usefulness}. The data collected from their interactions, as well as their responses to a post-execution questionnaire (included in the replication package), are comprehensively discussed, paving the way for future studies that consider the improvement of test generation tools' performance based on interactivity.
\end{itemize}

The paper is structured as follows. Section~\ref{sec:background} provides background on SBST, interactive SBSE and DynaMOSA (EvoSuite's default algorithm). Section~\ref{sec:approach} describes the design and implementation of \acronym. The experimental methodology is presented in Section~\ref{sec:study}. The results are detailed in Section~\ref{sec:results}, followed by a discussion in Section~\ref{sec:discussion}. Sections~\ref{sec:threats} and~\ref{sec:related} present threats to validity and related work, respectively. Section~\ref{sec:conclusion} concludes the paper.

%-----------------
% 2 - BACKGROUND
%-----------------

\section{Background}
\label{sec:background}
%-----------------
% 2.1 - SBST
%-----------------

The following two subsections present relevant concepts related to: 1) the generation of test cases through search-based techniques; 2) interactive approaches in SBSE and, more specifically, SBST; and 3) DynaMOSA, the default evolutionary algorithm in EvoSuite.

\subsection{Search-based test case generation}
\label{subsec:search-based}

Search-based software testing (SBST) has become a cost-effective approach to overcome the impracticality of exhaustive testing. Successfully applied in the context of test data generation~\cite{McMinn:2004:SST}, it has also been used to test object-oriented software~\cite{Wappler2006,fraser2012mutation}. Test suite generation can be formulated as a search problem, where test cases will be generated for all possible \emph{test coverage targets} given some coverage criteria, such as the lines of the system under test (SUT), its branches, or the injected mutations, i.e., seeded artificial faults in the code which are meant to be detected by the test suite. One of the most popular SBST tools is EvoSuite, which provides algorithms and objective formulations to generate test suites  for Java classes with a high fault-detection capability~\cite{fraser2012mutation}.

EvoSuite is based on evolutionary computation, a search approach in which a set of candidate solutions (a.k.a. individuals) is iteratively improved. In EvoSuite, such a population of solutions can be either whole test suites~\cite{Rojas2017-Whole} or individual test cases which are later grouped to create a final test suite~\cite{Panichella2018}. Test cases are represented as a sequence of statements with constructor and method calls, as well as their parameter values. An initial population of test cases is created by choosing such statements and values at random. In each iteration of the search (a.k.a generation), the evolution happens as follows: 1) a subset of individuals are selected for reproduction based on their quality; 2) individuals are modified by crossover and mutation operators, generating new individuals; and 3) the current population and the new individuals are compared to promote the survival of the best ones. A key element in any evolutionary algorithm is the fitness function, which assesses the quality of the individuals. In EvoSuite, such a function is configurable and maximizes one or more testing criteria. %~\cite{Gay2017}.
Additional evaluation criteria can be defined as secondary objectives, such as the test length to control the growth of the sequence of calls in the test cases. After the search, tests are completed with assertions based on mutation analysis, and a minimization step is executed to remove redundant constructs.

%-----------------
% 2.2 - ISBST
%-----------------

\subsection{Interactivity in SBST}
\label{subsec:isbse}

Interactive SBSE (iSBSE) promotes the active participation of software engineers by providing intermediate results for their inspection~\cite{Ramirez2019}. Their feedback is later integrated into the search process in order to progressively adapt it to the human's preferences. Designing an iSBSE proposal involves several decision factors that should be adapted to the application domain~\cite{Ramirez2019}. Firstly, the type of algorithm depends on the goal pursued by the interaction, e.g., including subjective evaluation criteria or refining the problem definition. Secondly, the user might be asked to perform different actions, such as the adjustment of the fitness evaluation, the selection and comparison of solutions, or the modification of the candidates. An interaction schedule has to be defined too, deciding the moment and frequency of interactions, and the criteria to choose the solutions to be shown. Finally, it is necessary to determine for how long the feedback influences the search, and whether such feedback can be modified.

Many software engineering tasks reformulated as optimization problems can benefit from the interaction with the engineer. Modeling and design involve complex cognitive tasks in which the engineer's subjective judgment becomes relevant to evaluate the quality of the candidates~\cite{Ramirez2018,Perez2022}. However, the use of interactive approaches has been broadened to other tasks that have been traditionally solved in a fully automated way, such as refactoring~\cite{Alizadeh2020,Rebai2022} and testing~\cite{Marculescu2015isbst,Marculescu2016,MarculescuFTP18}. Our prior work provides an overview of the possibilities that interactive optimization can bring to search-based test generation~\cite{Ramirez2021}. Based on the general design factors mentioned above~\cite{Ramirez2019}, we analyzed the suitability of several interactive options to address common limitations in search-based test generation, including potential actions a tester could do. In that work, we also discussed novel ideas related to the interaction scheduling, such as starting interaction after reaching a minimum coverage threshold or showing context information for the tester. The design of \acronym, which is presented in Section~\ref{sec:approach}, emerges from some of these ideas.

\subsection{Test-case generation with DynaMOSA}
\label{subsec:dynamosa}

DynaMOSA, proposed by Panichella et al.~\cite{Panichella2018}, is the state-of-the-art algorithm in the current version of EvoSuite. Since \acronym takes DynaMOSA as the base algorithm, we next summarize the characteristics relevant to our proposal.

\vspace{2pt}
\noindent\textbf{\emph{Target-oriented many-objective search.}}
DynaMOSA is a many-objective algorithm that evolves a population of individual test cases. It considers the individual distances from the targets in the class under test as the set of objectives to optimize. Hence, test cases are not evaluated by a single fitness function but by a set of $k$ functions, each one minimizing a different coverage target. Formally, the optimization problem is defined as finding a set of test cases $T = \{t_1,\ldots,t_n\}$ that minimize the fitness functions $f_1,\ldots,f_k$ associated with $k$ possible targets $U = \{u_1,\ldots,u_k\}$.

Each fitness function $f_{i}$ is defined as a distance, $d(u_i,t_j)$, which is different depending on the coverage criteria and expresses how close the execution traces of test case $t_j$ are from covering the target $u_i$ (see~\cite{Panichella2018} for details). The algorithm does not focus on all targets at the same time but dynamically chooses new targets as other targets at a higher position in the control dependency hierarchy are covered. 

\vspace{2pt}
\noindent\textbf{\emph{Preference criterion and selection.}}
DynaMOSA aspires to build test cases that fully cover the targets rather than test cases with acceptable values for all objectives. To guarantee this, DynaMOSA introduces a particular \emph{preference criterion} when comparing test cases. Namely, this is a target-oriented preference criterion because, given an uncovered test target $u_i$, a test case $t_1$ is preferred over another $t_2$ \emph{iff}:
\begin{equation}
\small{
    f_i(t_1) < f_i(t_2){~}OR{~}f_i(t_1) = f_i(t_2) \land length(t_1) < length(t_2)
    \label{eq:preference-criterion}
    }
\end{equation}

where $f_i(t_1)$ is the fitness value of test case $t_1$ for $u_i$, and $length$ is measured in terms of number of statements. Therefore, if two test cases have the same fitness for a given target, the shorter one is preferred. The survival strategy prioritizes test cases in the population which are closer to covering the uncovered targets currently addressed. If there is still room in the population after this, some of the remaining tests are selected based on a diversity preservation measure~\cite{Koppen2007}.

\vspace{2pt}
\noindent\textbf{\emph{Archiving.}}
Once the algorithm produces a test that covers one of the current uncovered targets, the efforts of the evolution are redirected towards the rest of the uncovered targets. To avoid losing useful test cases, they are archived together with the target they cover~\cite{Rojas2017-Whole}. If several test cases cover the same target, test length is used to discriminate. From now on, this archive will be named as \emph{coverage archive}. DynaMOSA uses it in two moments. Firstly, the reproduction step includes the mutation of test cases chosen from the archive with a certain probability. Secondly, the final test suite is built from the archive, whose test cases are post-processed to minimize them and add assertions. Notice that redundant test cases, i.e., those subsumed by other test cases when they cover the same targets, are removed at this stage.

%-----------------
% 3 - APPROACH
%-----------------

\section{$\acronym$: Incorporating interactive readability assessment}
\label{sec:approach}

A limitation in current SBST tools is that software engineers cannot easily understand the purpose of the generated test cases. To address this issue, these tools could implement mechanisms to involve the tester ---who has the knowledge of the class under test--- in the process. Here, we explain how our interactive approach, \acronym, allows testers to incorporate their subjective evaluation of test cases in terms of readability.
In the following subsections, we first present an overview of our interactive proposal and then we describe how we extended DynaMOSA to incorporate interactivity and support readability assessment.

\subsection{Overview of the interactive approach}
\label{sec:interactive-approach}

Before going into the technical details of \acronym, we first provide an overview of the approach by defining each of the general components of any interactive SBSE proposal. Such components need to take into account some particularities of how EvoSuite manages candidate solutions during the search. 
They also need to be tailored to the purpose of the interaction, i.e., readability assessment. 

\vspace{2pt}
\noindent\textbf{\emph{Type of algorithm.}}
Achieving high coverage is still the primary objective of any SBST approach. Therefore, our approach does not follow the precepts of human-based evaluation, where the fitness function is partially or totally replaced by a human. Instead, the extent to which a test case is readable ---as perceived by the tester--- will be used to compare test cases at certain moments of the search. This type of action is more aligned with a preference-based interactivity~\cite{Ramirez2019}, where the goal is to incorporate the user's choices to adapt the search process. At this point, it is important to distinguish between two different versions of the same test case:

\begin{itemize}
    \item \emph{Inner test case (individuals)}: This refers to the structure of the test cases internally handled and evolved by the evolutionary algorithm.
    \item \emph{Minimized test case}: This refers to the final appearance of a test case after minimization, i.e., once the statements not strictly required to meet the coverage criteria are removed. These are the test cases usually observed by testers at the end of the process.
\end{itemize}

Accordingly, in this approach, the tester will only review \emph{minimized versions} of some selected test cases. This implies that those inner test cases selected for interaction will be minimized before being shown to the tester.

\vspace{2pt}
\noindent\textbf{\emph{Type of feedback.}}
At some points during the execution, the search will pause and will enable an \emph{interaction moment}. In each one, the algorithm will dispatch one or more interactions to the tester, preparing some files for their inspection. A single \emph{interaction} consists of:

\begin{itemize}
    \item \emph{A coverage target} selected from the set of already covered targets. 
    \item \emph{A set of candidate test cases} ---in their minimized form--- covering the selected target. 
\end{itemize}

Focusing on a single target at a time allows the selection of candidate test cases with a common purpose ---and, therefore, similar tests that can be more easily confronted. Furthermore, the target itself provides a context that can help in the evaluation. Given that these candidates already cover the selected target, the tester will always review complete test cases (i.e., as they would appear in the final test suite) instead of partial ones and, thus, the tester can directly take care of their readability. Each candidate test case will store a value associated with its readability, called \emph{readability score}. It is the tester who has to assign these readability scores upon the revision and subjective assessment of each of the candidate test cases. \new{This is a numerical score within a range, whose limits can be configured. For example, a rating scale from 0 to 10 would be a valid range.}

\vspace{2pt}
\noindent\textbf{\emph{Interaction schedule.}}
Fatigue is an important issue to prevent in case of human interaction, as it could distort the process or even cause the tester to abort. Therefore, we should avoid situations requiring many or too frequent interactions, or the analysis of too much information. As such, the implemented interactive algorithm should present a configurable set of interaction-related parameters, so that testers can adapt the interaction to their needs. Essentially, the tester should be able to adjust (1) \emph{the number of test cases shown in each interaction}, (2) \emph{the number of interaction moments as well as the number of selected targets (single interactions) in each interaction moment}, and (3) \emph{when to start the interactions and their frequency}. Preparing two or more single interactions in one interaction moment can be useful to impose some constraints in the process of target selection, e.g., the selection of targets affecting different methods to foster diversity in the revision of test cases. This point will be further explained in Section~\ref{subsec:interaction-moments}.

\vspace{2pt}
\noindent\textbf{\emph{Information integration.}}
After the interaction, the search will resume and the feedback provided by the tester will be integrated according to the following strategy:

\begin{itemize}
    \item The most readable test cases for the tester will be saved and used to create the final test suite.
    \item All the assessed test cases will also be saved for later interactions. In this way, the system will avoid requesting the tester's opinion about minimizations already shown during the execution.
    \item The most readable test cases will be used as a source to generate new test cases in the population. This can help derive new test cases that preserve part of the characteristics that make them readable for the tester.
\end{itemize}

The final goal of these interactions is to reach a more readable test suite at the end of the process. Therefore, it is important to note that, beyond considering this new source to generate additional test cases, our interactive proposal does not interfere with the search and, thus, it is devised to preserve the coverage that would be achieved with the fully automated version.

\subsection{Interaction moments and target selection}
\label{subsec:interaction-moments}

\begin{figure}[ht]
\centering
\includegraphics[width=9cm]{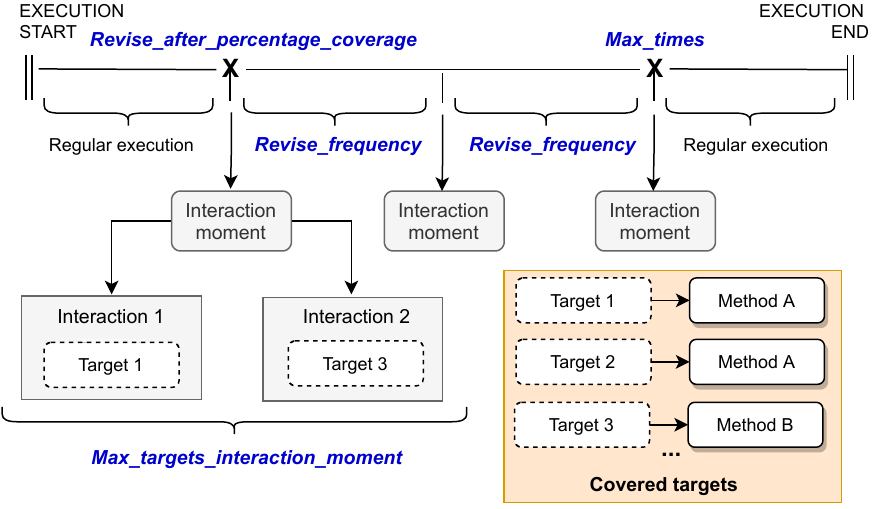}
\caption{General diagram of \acronym's interaction scheduling and target selection.}
\label{fig:interaction-diagram}
\end{figure}

\acronym modifies the original routine of DynaMOSA by pausing the search and enabling certain interaction moments during the execution to allow testers to assess the readability of some generated test cases. These interactive moments will be scheduled at a given \emph{Revise\_Frequency} ---a configurable parameter---, and will take place at that frequency until exhausting the maximum number of interactions indicated by the parameter \emph{Max\_times}. Our version also includes a novel option to customize when interactivity should start: \emph{Revise\_after\_percentage\_coverage}. More specifically, the first interaction moment is only enabled once a certain percentage of the global test coverage is reached. This way, testers can decide whether they are willing to review test cases at an early stage, or prefer to wait until a greater number of targets have been covered. The process is graphically explained in Figure~\ref{fig:interaction-diagram}.

\begin{table*}
\renewcommand{\arraystretch}{1.1}
\caption{Definition of interaction parameters in \acronym.}
\label{tab:schedule}
\vspace{-0.3cm}
\centering
\scalebox{0.95}{
\begin{tabular}{|l|l|l|}
\hline
\textbf{Category} & \textbf{Parameter} & \textbf{Description} \\
\hline
\textbf{Adjustment}       & \emph{Revise\_frequency}         & Interaction will take place at regular intervals of this number of generations.\\
\textbf{of interaction}   & \emph{Max\_times} & Maximum number of times the user is willing to interact during the search. \\
\textbf{time}             &  \emph{Revise\_after\_percentage\_coverage} & The possibility to interact will be enabled when reaching this percentage of coverage.\\
                          & \emph{Max\_targets\_interaction\_moment }   & Maximum number of targets that the user is willing to address in one interaction moment. \\ 
\hline
\textbf{Solutions shown}  & \emph{Percentage\_to\_revise} & Percentage of test cases in the population the user is willing to revise at most. \\
\hline
\textbf{Readability score} &  \emph{Max\_readability\_score} & Allowed readability scores in the range [0, Max\_readability\_score].   \\
\textbf{and}              & \emph{Readability\_threshold} & Readability score under which test cases will not be transferred to the preference archive.\\
\textbf{preference archive} & \emph{P\_preference\_selection} & Probability of selecting a test case from the preference archive to breed a new test case.\\
\hline
\end{tabular}
}
\end{table*}

At an interaction moment, one or more single interactions can be prepared, depending on the parameter \emph{Max\_targets\_interaction\_moment} (2 in the example). An interaction starts by extracting one target from the list of already covered targets. When this parameter is set to a number greater than 1, an interaction can only select targets affecting methods other than those previously addressed in the same interaction moment. Looking at the example, target 1 and 2 are related to method A, and target 3 to method B. Since the first interaction already focused on target 1 (method A), the second interaction discards target 2 ---because it also has to do with method A--- and takes target 3 (method B) instead. This constraint of selecting different methods seeks to diversify the tests shown so that the revision encompasses a greater part of the code of the class under test. 

As mentioned above, the target provides a context for the comparison of test cases ---as they have in common that they cover the same target. Therefore, to help understand the target, the output format of different coverage criteria has been restructured for pretty-printing the target to a file. Table~\ref{tab:schedule} collects and describes the parameters mentioned in this section under the category of ``adjustment of interaction time'' (following the terminology by Ramirez et al.~\cite{Ramirez2019}). Finally, note that the strategy to select the targets for interactions will be subject to evaluation later on in Section~\ref{sec:study}.

\subsection{Selection of candidate test cases}
\label{subsec:candidates}

In each interaction, a number of candidate test cases is selected for the coverage target. The test cases to be shown are primarily extracted from the population. Even though the search evolves the test cases for uncovered targets, some of them can still cover the selected target indirectly. Note also that an inner test case can transform into different minimized versions, depending on the target driving the minimization process. This is because the unnecessary statements removed in the minimization can be different depending on the portion of the code pointed by the target. Therefore, the same inner test case can give rise to different minimized candidates for different interactions.
The percentage of tests to review is defined as a parameter (\emph{Percentage\_to\_revise}) so that the tester can control the amount of information requested (see Table~\ref{tab:schedule}, category ``solutions shown''). Therefore, the maximum number of tests cases to revise is computed as follows: 
\begin{equation}
\small{
    NT = \textnormal{max}({~}Population \times Percentage\_to\_revise,\quad 2)
    \label{eq:nt-calculation}
    }
\end{equation}

where \emph{Population} represents the size of the set of test cases evolved in the search population, and 2 is the minimum imposed so that the user can have at least two candidates to compare. The test case associated with that target in the coverage archive occupies a spot on the set of candidate test cases of size $NT$. Then, if the number of individuals in the population covering the target is greater than $NT-1$, then $NT-1$ test cases are selected at random.

\subsection{New archives: preference and readability} 
\label{subsec:new-archives}

In \acronym, the improvement of the readability of the final test suite according to the tester's preferences is considered as an additional objective to the coverage of test targets. Therefore, apart from the aforementioned \emph{coverage archive}, the implemented interactive version introduces two new archives to handle this complementary objective: the \emph{preference archive} and the \emph{readability archive}.

\vspace{2pt}
\noindent\textbf{\emph{Preference archive.}}
This archive keeps track of the most readable test case for each of the targets for which the tester interacts. More specifically, the preference archive saves a list of tuples with the following information: $<$\texttt{target}, \texttt{minimized{~}test}, \texttt{readability{~}score}$>$. This archive is maintained because of the following three reasons:
\begin{itemize}
    \item To keep testers informed about the progress/to let them know that their opinion is taken into account.
    \item To help form the final test suite (see Section~\ref{subsec:final-test-suite}).
    \item To take part in the production of candidates in future generations.
\end{itemize}

Since this archive records the most readable test cases, it forms a valuable source of information (see ``Information integration'' in Section~\ref{sec:interactive-approach}). To exploit this opportunity, we include a new parameter called $P\_preference\_selection$. This parameter indicates the probability for the algorithm to select the preference archive instead of the coverage archive as a source for the generation of new tests (see ``Archiving'' in Section~\ref{subsec:dynamosa}). Note that this parameter comes into play once the preference archive is populated with some test cases as a result of the first interactions.

\vspace{2pt}
\noindent\textbf{\emph{Readability archive.}}
Minimized test cases whose readability has been already valued are saved in this archive together with their readability score. Right before a newly scheduled interaction, the archive is queried to avoid requesting the tester's opinion about those minimizations again. If a minimized test case is found in the readability archive, that test case directly receives the recorded score for that minimization. Additionally, the test case and its associated score are shown to the tester so that it can be observed and used as a reference when evaluating test cases not yet seen. 

\begin{figure}[ht]
\centering
\includegraphics[width=8.8cm]{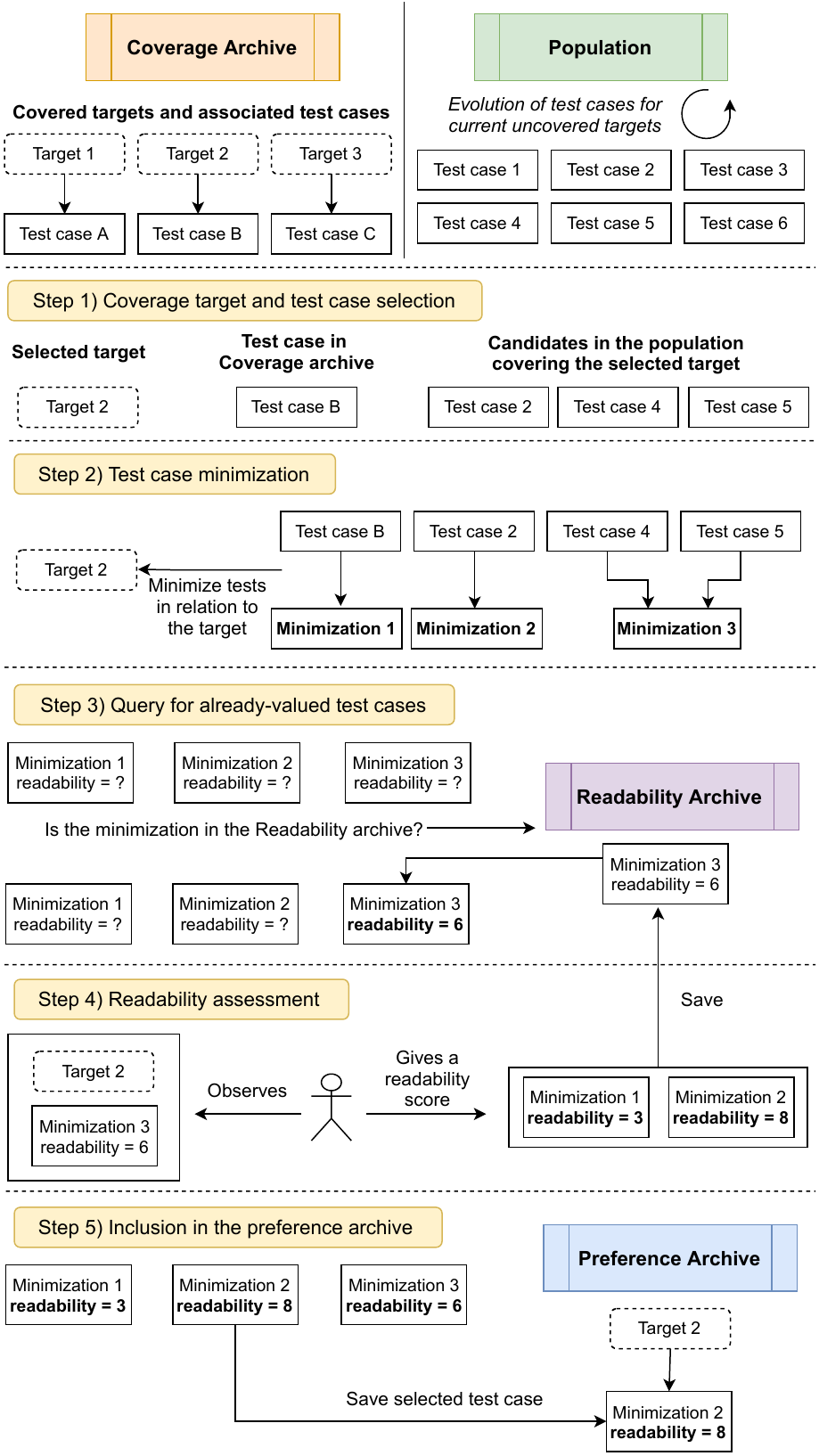}
\caption{General workflow of the process of readability integration.}
\label{fig:readablility-incorporation}
\end{figure}

\subsection{Integration of readability scores}
\label{subsec:incorporation}

Figure~\ref{fig:readablility-incorporation} depicts a workflow and an example of the process of readability integration when an interaction is scheduled.  

\vspace{2pt}
\noindent\emph{Initial state.} Initially, the coverage archive contains those targets already covered (3 targets in the example) and their respective covering test case, whereas the population consists of the tests (6 in the example) evolved in relation to the uncovered targets currently addressed by the algorithm.

\vspace{2pt}
\noindent\emph{Step 1: Coverage target and test case selection.} When an interaction moment is enabled, a target is first selected from the coverage archive (target 2 in the example). At that moment, the algorithm starts a search of candidate test cases for the interaction. The first candidate is the test associated with the target in the archive (test case B). The rest of the candidates are extracted from the population. In this example, we assume that test cases 2, 4 and 5 also cover target 2. 

\vspace{2pt}
\noindent\emph{Step 2: Test case minimization.} The candidates are subject to the minimization process for the selected target. As a result, some of the test cases can result in the same minimization (this is the case of test cases 4 and 5), thus reducing the number of candidates (from 4 test cases to 3 minimizations).

\vspace{2pt}
\noindent\emph{Step 3: Query for already-valued test cases.} Except for the first interaction, the readability archive will normally contain some minimized tests valued in previous interactions. The readability score of those minimized candidates already included in the archive is retrieved (minimization 3 is assigned 6 as its readability score). Depending on the effect of the archive, the tester may need to evaluate a number of test cases between 1 and $NT$. As the last step, assertions are also added to the minimized test cases if the option is enabled.

\vspace{2pt}
\noindent\emph{Step 4: Readability assessment.} In this step, minimized test cases are shown to the tester for revision and readability evaluation. The coverage target and already-valued minimizations are also shown and serve as a reference point for the assessment of not yet seen tests. To this end, the system outputs the corresponding files with the information of the test cases and the target for their revision. The system is paused during the revision and only resumes the execution once the tester provides the required readability scores through the console. Valid scores range from 0 to $Max\_readability\_score$, which can be adjusted. As mentioned earlier, once the minimized test cases are scored by the tester, they are saved in the readability archive. 

\vspace{2pt}
\noindent\emph{Step 5: Inclusion in the preference archive.} The test case with the best readability score is transferred to the preference archive---in its minimized version---associated with the selected target (minimization 2). In case of a draw, one of the tied candidates is selected at random. In contrast, if all candidates had been valued with a score under \emph{Readability\_threshold}, none of them will be transferred to the archive.

This 5-step process is repeated in the same interaction moment up to \emph{Max\_targets\_interaction\_moment} times or until \emph{Max\_times} interactions are completed. Finally, note that, at the end of an interaction moment, the user is informed of any update of the preference archive. In fact, the system outputs the preference archive and the tester can consult it to know about the progress of their preferences. All mentioned interactive parameters related to test readability and the preference archive are shown in a special category in Table~\ref{tab:schedule}.

\vspace{2pt}
\noindent\emph{Exceptional cases.} It is worthwhile mentioning some special cases that can alter this process. A few situations may arise that make the interaction unnecessary for a given target:

\begin{itemize}
    \item \emph{Step 1}: The target is not covered by any of the test cases in the population; therefore, the number of candidates is lower than $NT$ (see Equation~\ref{eq:nt-calculation}). 
    \item \emph{Step 2}: All selected inner test cases derive in the same minimized version (again, the number of candidates would be lower than $NT$).
    \item \emph{Step 3}: All minimized versions are directly found in the readability archive. 
\end{itemize}

In these three cases, the algorithm selects a new target and repeats the process of candidate selection. If the list of covered targets was exhausted, interactions at that moment would be skipped and the execution would resume. 

Another special case is that the same target can be addressed in different interaction moments during the execution. The rationale behind this is that new candidate test cases may be available in the evolved population whose minimization is different from those previously observed.
If the selected target appeared in a past interaction, the preference archive may contain a test case for that target. This possibility has two implications:

\begin{itemize}
    \item \emph{Step 1:} Instead of two or more candidates, \acronym is allowed to progress with the interaction even if only one candidate is found in the population because this test case can be compared with the one already saved in the preference archive.
    \item \emph{Step 5:} The preference archive is only updated when the readability score of the new candidate is higher than the score of the test case saved in the preference archive for that target. If both have the same score, the shortest test case will be preferred.
\end{itemize}

\subsection{Final test suite}
\label{subsec:final-test-suite}

At the end of the execution, the final test suite is created by aggregating archived test cases generated during the search.
Unlike the original version, the priority source of test cases in \acronym is the preference archive instead of the coverage archive. Thus, all the tests in the preference archive are added to the test suite. However, in case that some redundant test cases are found, those with the lowest readability score are removed first, keeping the most readable tests at all times. Since testers only address a subset of all targets, the test suite is complemented with the coverage archive, adding all the test cases required to maintain the same coverage level reached at the end of the search.

%-----------------
% 4 - Methodology
%-----------------

\section{Experimental methodology}
\label{sec:study}

This section details the methodology followed to conduct the experimental evaluation of \acronym. Firstly, the research questions (RQ) are stated. Then, we describe the scope and analysis method of a first experiment where we seek to establish a suitable target selection strategy for the case of interactive readability assessment. Finally, we present the characteristics of the interactive experiment with humans using \acronym, including the selection of participants, survey design and interaction parameters.

\subsection{Research questions}
\label{subsec:rqs}

In the first part of our experiments, we aim at determining a proper target selection strategy for the interactions. It would be interesting to prioritize those covered targets requiring more elaborated test cases---more interesting from the perspective of readability assessment---to make most of the tester's revision. At this point, we wonder whether those targets covered at a later stage during the execution---seemingly harder to cover---actually require larger test cases. The rationale is that the longer the test case, the more opportunities to have test cases with different code lines. This leads to the following research question (RQ):

\begin{itemize}[leftmargin=*]
\item[] \textbf{RQ1:} \emph{What is the relation between the targets covered throughout the search and the length of the test cases covering them?}
\end{itemize}

Once the test selection strategy is decided, the second part of our experimentation involves participants who interact with \acronym. Namely, we want to know about three different aspects: \emph{test readability}, \emph{interaction process} and \emph{usefulness} of the approach. The motivations of this study are reflected in the following research questions:

\begin{itemize}[leftmargin=*]
    \item[] \textbf{RQ2:} \emph{How do the participants address the evaluation of test case readability?}
    \item[] \textbf{RQ3:} \emph{How do the participants interact with \acronym?}
    \item[] \textbf{RQ4:} \emph{\new{Do the participants find \acronym useful?}}
\end{itemize}

\subsection{Description of Experiment \#1}
\label{subsec:exp-procedure-rq1}

The first experiment analyses the relation between covered targets and test length (RQ1). To this end, we instrumented EvoSuite to: 1) record the generation when DynaMOSA covers each target, and 2) minimize the test cases contained in the coverage archive as a further step at the end of the execution. For each of the different minimizations, a file then collects the generation in which the test case was inserted in the archive and the number of lines and characters of the minimized test case (assertions are not included in the counting because they are only added once the target has been covered). Since the minimization of many test cases may take significant time, this step is only considered for experimental purposes and not for the target selection process in the interactions. To make this analysis affordable and simplify the reporting of results, we used 10 classes of a varying number of targets from the \emph{SF110} corpus\footnote{SF110 Corpus: \url{https://www.evosuite.org/experimental-data/sf110/} (Last accessed: 21st September 2022)} and the EvoSuite's tutorial. This set also includes the class \texttt{ArrayIntList}, used in the second experiment. We used \emph{Branch Coverage}, \emph{Line Coverage} and \emph{Weak Mutation} as the set of coverage criteria to be optimized to keep the same conditions with respect to the interactive experiment, for which a reduced number of criteria was preferred to avoid information overload (see Section~\ref{subsec:params}). Based on the guidelines for assessing algorithms of random nature~\cite{Arcuri2014}, we executed DynaMOSA 30 times for each class. The rest of the search parameters are configured to the default values.

To analyze the results, we collect the records of the 30 executions and then divide them in 3 different groups according to the moment when the tests were inserted into the archive: initial population (\emph{generation 0} or \emph{g0}), between the \emph{generations 1 and 9} (\emph{g1-9}), and at \emph{generation 10 or later} (\emph{g10+}). The number of available targets to cover decreases as the execution progresses, so this division seeks to balance the number of records in each group. Then, for each group, we calculate the average test length in terms of lines and characters. We also obtain the minimum values, as we are especially interested in preventing interactions with too simple tests. Wilcoxon and effect size (Cliff's delta) tests are applied to statistically analyze the results with $\alpha=0.05$. \new{The data distributions under analysis correspond to average test length (lines or characters) measured at different moments of the search. Therefore, the null hypothesis establishes that the length remains the same (on average) during the search.}

\subsection{Description of Experiment \#2}
\label{subsec:exp-procedure-rq2-4}

The second experiment, in which participants interact with \acronym assessing and providing scores for some automatically-generated test cases at some points during the search, is aligned with RQ2-RQ4. First, we detail the organization of the experiment, the selection of participants and the data collection process. Next, we focus on the configuration of general and interactive parameters.

\subsubsection{Study design}

\vspace{2pt}
\noindent\textbf{\emph{Organization.}} The experiment was planned as a session scheduled in three parts: 1) a short introduction to the context of the experiment, in which we explain basic concepts about unit testing and EvoSuite, as well as the task to be done; 2) a running example of \acronym using the Eclipse IDE; and 3) the actual execution of the algorithm, independently done by each participant, followed by some time to fill a survey. We considered steps 1 and 2 to be necessary since participants might not have experience in the automated generation of test cases, and they also need to understand concepts like the coverage criteria and targets used by \acronym. Due to the inherent fatigue that an interactive experiment imposes, we carefully designed the interaction scheme (details are provided in Section~\ref{subsec:params}) and the experimental conditions. In total, the experiment was conceived to last a maximum of 2 hours, setting 10 minutes to get familiar with the class under test and \new{a maximum of} 40 minutes for the execution of \acronym in step 3. 

For the running example (step 2), we made use of a simple \texttt{Stack} class to explain how the interaction happens and to allow participants to get familiar with the environment. For step 3, the actual interactive experiment, we chose \texttt{ArrayIntList} as the class under test. This is a well-known class of medium complexity used in other SBST experiments with participants related to test readability improvement~\cite{Panichella2016, Roy2020}. Due to the time required to complete the experiment and the cognitive burden of manually revising many tests, all participants used one class under test. Actually, the same class was used to be able to compare their results. However, each participant used a different random seed for the execution, meaning that each one evaluates different test cases. \new{A step-by-step example of the interactions of one participant can be found here\footnote{\url{https://github.com/PdedP/InterEvo-TR/tree/master/Example of interaction}}.}

\vspace{2pt}
\noindent\textbf{\emph{Participants.}} Participants were recruited by email, sending invitations to department members of the two universities involved in the study. We extended the invitations to professional developers among our industrial contacts, as well as student collaborators. In total, we obtained positive answers from \new{39} participants (similar to other studies~\cite{Panichella2016, Roy2020}), divided into: 4 Bachelor's students, 4 Master's students, 7 PhD students, 10 faculty members with programming skills and \new{14} professional developers from industry. As part of the survey, we collected some basic information about their experience programming in Java and testing software in general. As shown in Table~\ref{tab:participants}, the majority of the participants have 3 or more years of experience coding in Java (\new{51\%}). Profiles with expertise in testing are more difficult to find as it requires more specialization. Nevertheless, \new{62\%} of the participants have more than one year of experience in testing. Among those with less than one year of experience, we mostly found students at different levels. We also asked participants to rate (1: low, 5: high) their own programming skills in Java, for which most of them (\new{64\%}) considered they have good skills (3 or more on the scale 1-5).

\begin{table}
\renewcommand{\arraystretch}{1.1}
\caption{Experience and skills of the participants.}
\label{tab:participants}
\vspace{-0.3cm}
\centering
\scalebox{0.95}{
\begin{tabular}{|l|r|r|r|r|r|}
\hline
\new{\textbf{Years of experience}} & \textbf{$<$1 \new{year}} & \textbf{1-2} & \textbf{3-6} & \textbf{7-10} & \textbf{$>$10} \\ \hline
Programming in Java & \new{10} & \new{9} & 12 & 3 & \new{5}\\
Testing software & \new{15} & \new{10} & \new{2} & \new{9} & \new{3}\\
\hline
\hline
\new{\textbf{Self-assessment}} & \new{\textbf{1-Low}} & \textbf{2} & \textbf{3} & \textbf{4} & \new{\textbf{5-High}} \\ \hline
Programming skills in Java & \new{5} & \new{9} & \new{13} & 10 & 2\\
\hline
\end{tabular}
}
\end{table}

\vspace{2pt}
\noindent\textbf{\emph{Questionnaire and data collection.}} 
The survey includes a number of predefined questions to be filled after completing the task, grouped into six parts: 1) test generation process, 2) test readability assessment, 3) aspects influencing their subjective scores, 4) interaction process, 5) perceived usefulness and 6) actions they would have liked to do. Table~\ref{tab:questions} shows the specific questions aligned to RQ2, RQ3 and RQ4. To avoid neutral answers, participants express their opinion using a 4-point Likert scale: Fully disagree, partially disagree, partially agree, fully agree. Finally, the survey provides a text area for open comments and suggestions. 

As for the perception of usefulness, participants had to compare the final test suite with two other test suites automatically generated with EvoSuite under the same parameter setting. This set did not include the test suite automatically generated with the same seed used in the interactive execution; by doing this, we try to prevent participants from finding in the test suite test cases poorly valued by them in the interactions, which could condition their perception of the results. 
\new{In a second part of this evaluation, we wanted to analyze whether participants perceived differences in the tests selected by the tools. Given the interactive nature of \acronym, participants cannot perform a ``blind'' execution of each tool. To address this issue, we asked them to compare individual test cases selected by both \acronym and EvoSuite in a posterior evaluation. More specifically, they had to assign a readability score for 3 different pairs of test cases: (1) \emph{interactive vs interactive}, two tests whose readability differed significantly for one of the participants in his/her execution of \acronym; (2) \emph{interactive vs automatic}, one test highly ranked by one participant in the execution of \acronym against another similar test automatically selected by EvoSuite to create the final test suite; and (3) \emph{automatic vs automatic}, two test cases selected from test suites automatically generated by EvoSuite, which acts as a control case. All the details about the methodology to carry out this study can be found in an extended report\footnote{\url{https://github.com/PdedP/InterEvo-TR/tree/master/Analysis/RQ2-RQ4/Extended technical report}}}.

The qualitative analysis of the survey responses is complemented with additional information saved in several log files: generations at which interactions happen, the number of selected test cases, the number of resulting minimizations and how many of those are found in the readability archive, the selected targets, the provided scores and the time spent in each interaction. The tests selected for each interaction, the preference archive and the final test suite are saved as Java files. The survey, the anonymized responses and execution results are available in the replication package.

\subsubsection{Parameter setting and interaction scheme}
\label{subsec:params}

\vspace{2pt}
\noindent\textbf{\emph{General parameter setting.}}
Most participants are not acquainted with the coverage criteria handled by EvoSuite, so dealing with too many coverage criteria could be overwhelming for them. This led us to limit the set of criteria to \emph{Branch Coverage}, \emph{Line Coverage} and \emph{Weak Mutation}, widely used in relevant experimental studies~\cite{Rojas2017-Whole}. During the interaction, information about the target is dumped into a text file. To make it more accessible to participants, the information provided in case of weak mutation includes the kind of mutation but not the exact change. Moreover, mutations are injected at the bytecode level and the change performed can be misleading sometimes, as it does not directly correspond with the source code. We set a search budget of 1,000 generations\new{, which is the approximate average number of generations computed after executing EvoSuite 30 times on this class with a timeout of 2 minutes.} Using generations instead of time as stopping condition allows us to control the frequency at which interactions are dispatched, as described in the following paragraph. We also set a global timeout of 40 minutes for the whole execution. The rest of the parameters were left with their default values. \new{Therefore, the initial population is initialized in the same way as EvoSuite does it by default (no new elements are added that could bias the readability results), and the tests observed by the tester are previously minimized and augmented with asserts.}

\vspace{2pt}
\noindent\textbf{\emph{Interaction parameters.}}
We adjusted the interaction parameters (see Table~\ref{tab:schedule}) based on the analysis of preliminary executions to control the cognitive load required by interactions\new{, and also on the basis of a previous study to observe the influence of different parameter settings in the results. For further information on this, the study can be found in the extended report of the replication package (see footnote)}. 
In this regard, we reached a consensus to set \emph{Max\_times} = 10 interactions with a maximum of 4 tests in each (\emph{Percentage\_to\_revise} = 8\% and population size = 50, therefore, \emph{NT} = 4). We also agreed to set \emph{Max\_targets\_interaction\_moment}~=~3 to promote the appearance of targets related to at least 3 different methods. 
With this configuration, the number of interaction moments varies between a minimum of 4 and a maximum of 6. This maximum could be reached if the system does not find targets to interact with in some of the previous interaction moments. This was achieved by dividing the search budget (in generations) into 5 equal segments and setting the result as \emph{Revise\_frequency}. Based on the analysis of RQ1, we configured the EvoSuite parameter \emph{Enable\_secondary\_objective\_after} = 10 to delay the activation of the first interaction moment (at least until reaching the $10^{th}$ generation), thus avoiding interactions focused on too simple targets. This parameter value, together with \emph{Revise\_after\_percentage\_coverage} = 50\%, should be reached before enabling the interactive process for the first time. Participants can rate each test case on a scale from 0 to 10 (\emph{Max\_readability\_score}), being \emph{Readability\_threshold} = 3. We opted for a wide range of values to observe whether participants made use of the full range of available values or just concentrated on a subset. Finally, we set \emph{P\_preference\_selection} = 20\%, meaning that 1 in 5 times the system will select the preference archive instead of the coverage archive to create a new test case \new{(in general terms, this means that 1 out of 50 test cases in the population are generated in this way in each generation on average)}.

\begin{table}[ht!]
\renewcommand{\arraystretch}{1.1}
\caption{Questions in the survey related to RQ2, RQ3 and RQ4.}
\label{tab:questions}
\vspace{-0.3cm}
\centering
\scalebox{0.92}{
\begin{tabular}{ll}
\hline
\textbf{ID} & \textbf{Question} \\
\hline
Q2.1 & The revised tests cases were easy to understand\\
Q2.2 & Could you appreciate differences regarding the readability of\\
     & test cases revised in the same interaction?\\
Q2.3 & Was it easy to assign a readability score to each of the test cases?\\
Q2.4 & Was the target useful to guide you when evaluating \\
     & the readability of the test cases?\\
\hline
Q3.1 & The number of interactions was appropriate\\
Q3.2& The number of tests to revise in each interaction was appropriate\\
Q3.3& The interactions were tedious and costly in time\\
Q3.4 & I paid more attention to earlier interactions than to later ones\\
Q3.5 & It was useful to be informed about the score assigned to\\
     & already revised test cases \\
Q3.6 & The preference archive was useful to know that my decisions\\
     & were taken into account and had an impact on the final result\\
\hline
Q4.1 & The review and comparison of tests helped me better under-\\
     & stand the purpose of some of the tests in the final test suite.\\
Q4.2 & The final test suite is more readable than the other\\
     & test suites provided for the same class\\
Q4.3 & The time spent on interactions has compensated for the result\\
\hline
\end{tabular}
}
\end{table}

%-----------------
% 5 - RESULTS
%-----------------

\section{Results}
\label{sec:results}

The following subsections present the results for our RQs.

\subsection{RQ1: Covered targets and test case length}
\label{subseq:rq1}

RQ1 studies the relation between the moment when targets are covered and the length of the test cases covering them.
Table~\ref{tab:results-length} shows the mean and minimum number of lines and characters of the minimized test cases for the covered targets in each class, divided by the 3 groups indicated in Section~\ref{subsec:exp-procedure-rq1}. The results are shown in ascending order of the number of targets in each class (\#T.). Test cases in g0 ---those covering targets before the evolution has even begun--- are on average of shorter length than those in the two other groups (g1-9 and g10+), both in terms of lines and characters in almost all classes. We can highlight the case of \texttt{CoordSystemUtilities}, where the difference between the mean in g0 and g10+ is 5.3 lines (9.1 vs 14.4) and 235 characters (474 vs 709). With the analysis of the minimum values, we wanted to prevent selecting too simple test cases for the interactions. As expected, seemingly trivial test cases appearing in the initial population (just with 1 line and less than 60 characters) are enough to cover some targets in 8 classes. However, and more interestingly, we cannot find such short test cases from that point on (except for \texttt{DateUtil} in the range 1-9). The case of \texttt{Player} is notable, where the shortest test case in g10+ has 12 lines and almost 500 characters, in contrast to the simplest test in g0, with 2 lines and 62 characters. The difference between the groups g1-9 and g10+ is not so marked, albeit the general tendency is the appearance of longer tests from the $10^{th}$ generation onward. This is especially evident when analyzing the minimum test case length in each group.

\begin{table}[ht!]
\renewcommand{\arraystretch}{1.1}
\caption{Relation between targets and length of the test cases covering them.}
\label{tab:results-length}
\vspace{-0.3cm}
\centering
\scalebox{0.95}{
\begin{tabular}{lclrrrrrr}
\hline
\multirow{2}{*}{\textbf{Class}} & \multirow{2}{*}{\textbf{\#T.}} & \multirow{2}{*}{\textbf{Stat.}} & \multicolumn{3}{c}{\textbf{\# Lines}} & \multicolumn{3}{c}{\textbf{\# Characters}}\\
%\cmidrule(l){4-6} \cmidrule(l){7-9}
 &  &  & \textbf{0} & \textbf{1-9} & \textbf{10+} & \textbf{0} & \textbf{1-9} & \textbf{10+} \\ \hline
Label & \multirow{2}{*}{70} & Mean & 4.5 & 7.7 & 6.8 & 132 & 232 & 201 \\
(\emph{jipa})                       &   & Min. & 1 & 2 & 2 & 24 & 56 & 56 \\\hline
ATM   & \multirow{2}{*}{87} & Mean & 7 & 11 & 11.5 & 259 & 415 & 434 \\
(\emph{tutorial})                       &   & Min. & 1 & 8 & 9 & 33 & 320 & 329  \\\hline
JSJshopNode & \multirow{2}{*}{146}  & Mean & 7.5 & 10.1 & 9.4 & 243 & 352 & 348 \\
(\emph{shop})                       &   & Min. & 1 & 1 & 3 & 46 & 78 & 129 \\\hline
XMLUtil & \multirow{2}{*}{232} & Mean & 5.2 & 5.6 & 6.8 & 237 & 277 & 318 \\
(\emph{checkstyle})                       &   & Min. & 1 & 2 & 2 & 25 & 72 & 187 \\\hline
bcWord & \multirow{2}{*}{295} & Mean & 4.5 & 5 & 5.3 & 196 & 233 & 246 \\
(\emph{battlecry})                       &   & Min. & 1 & 3 & 4 & 59 & 119 & 190 \\\hline
ArrayIntList & \multirow{2}{*}{403} & Mean & 5.4 & 5.1 & 5.7 & 195 & 192 & 209 \\
(\emph{commons p.})                      &    & Min. & 1 & 2 & 3 & 53 & 80 & 123 \\\hline
Player & \multirow{2}{*}{468} & Mean & 9.5 & 12 & 14.5 & 384 & 485 & 571  \\
(\emph{gangup})                        & & Min. & 2 & 4 & 12 & 62 & 145 & 498 \\\hline
User  & \multirow{2}{*}{657} & Mean & 5.8 & 7.2 & 7 & 221 & 278 & 273 \\
(\emph{lhamacaw})                        & & Min. & 1 & 2 & 1 & 26 & 67 & 37 \\\hline
CoordSystem. & \multirow{2}{*}{824} & Mean & 9.1 & 11.1 & 14.4 & 474 & 563 & 709 \\
(\emph{jopenchart})                        & & Min. & 1 & 4 & 4 & 182 & 258 & 258  \\\hline
DateUtil & \multirow{2}{*}{1363} & Mean & 3.2 & 3.9 & 3.6 & 117 & 140 & 124 \\
(\emph{caloriecount})                        & & Min. & 1 & 1 & 1 & 32 & 31 & 35 \\
\hline
\end{tabular}
}
\end{table}

Given these results, we run the Wilcoxon test to know whether the observed differences between g0 and the other two groups are statistically significant. To that end, we join the results of g1-9 and g10+ in a single group. Therefore, the null hypothesis establishes whether the number of lines (equivalent for characters) in the test cases from the initial population (g0) are similar to those generated afterwards (g1+). The test indicates that statistical differences exist for all classes, both in terms of lines and characters. The exception is \texttt{ArrayIntList}, where the average test in g1-9 was already slightly shorter than that in g0. According to the effect size test, the differences are \emph{large} in the rest of the classes, except for the pairs $<$\texttt{XMLUtil}, lines$>$ (\emph{small}) and $<$\texttt{DateUtil}, characters$>$ (\emph{medium}).

\begin{center}
\begin{tcolorbox}
  [colback=white,colframe=gray!75!black,fonttitle=\bfseries,width=9cm]
\textbf{RQ1 answer:} \emph{Targets covered at the beginning require of much shorter test cases than the rest of targets covered during the search. Focusing on the minimum size, the set of tests created in the search do not include as short test cases as those appearing in the initial population in general.} 
\end{tcolorbox}
\end{center}

We are now in a position to determine which targets are worth selecting first, especially because the number of interactions is limited. Thus, prioritizing targets not covered directly from the beginning seems a reasonably good strategy to increase the chances of inspecting more elaborated tests. Consequently, the target selection strategy for the experiment with participants prioritizes the most recently covered targets. Furthermore, in the case of \texttt{ArrayIntList}, test cases for targets covered later (g10+) are often longer than those covered previously and, more importantly, they are never too short (at least 3 lines and 123 characters).

\begin{figure*}[ht]
\centering
\includegraphics[width=0.95\textwidth]{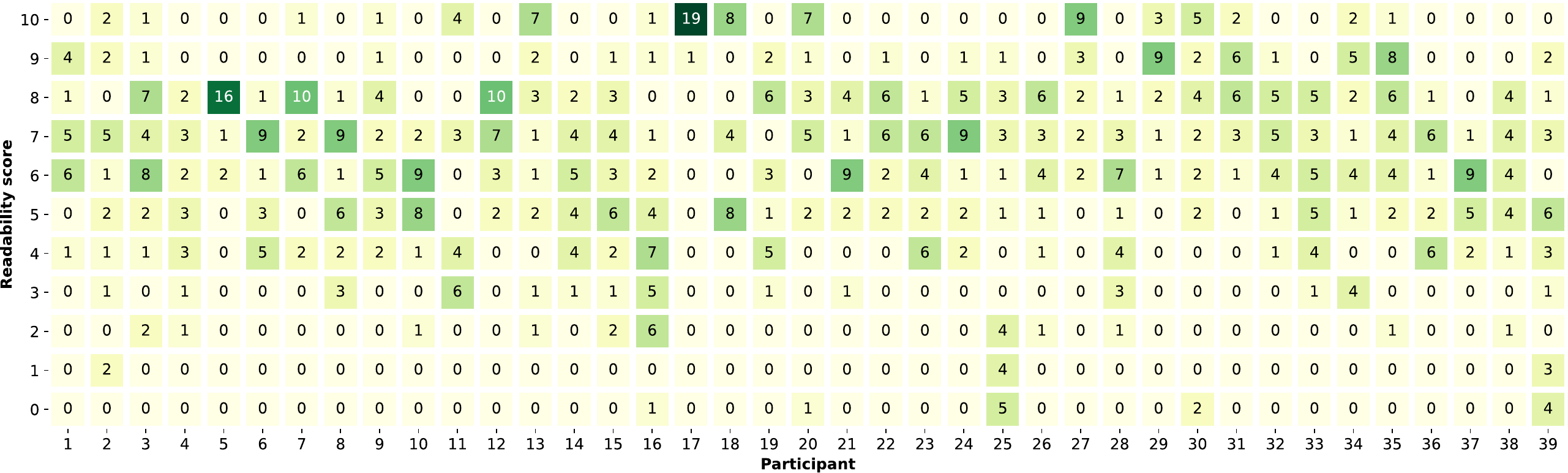}
\caption{Frequency of readability scores (0-10) assigned by each participant during the 10 interactions.}
\label{fig:readability-score-frequency}
\end{figure*}

\subsection{RQ2: Test readability}
\label{subseq:rq2}

In RQ2, we seek to analyze the behavior of the participants regarding the interactive readability assessment process. Since each participant rated different test cases for different targets at different moments, we manage their readability scores independently. Figure~\ref{fig:readability-score-frequency} shows how frequently each score was assigned. i.e., how many times the readability score appearing in the vertical axis was selected. As can be observed, only \new{5} participants assigned the minimum score (0) at some moment, whereas scores 1 and 2 were chosen by \new{33}\% of participants at least once. In total, only \new{5.7}\% of all tests received scores lower than 3 (value of \emph{Readability\_threshold}), suggesting that they did not consider automatically-generated test code for this class as particularly difficult to understand. Most of the participants used a wide range of values, concentrating on scores between 5 and 8. We also detect a couple of students (participants with id 5 and 17) who perceived most of the tests as highly readable, repeatedly assigning high scores (8 and 10, respectively).\footnote{\new{Both participants completed the task very quickly, and their responses to the survey reveal less critical thinking than the rest of participants. Their unusual behavior seem to indicate they found the task easier than expected, and they did not engage too much.}} Attaching the same score to all the test cases shown in a given interaction was not unusual, it happened \new{25.4}\% of the time they revised at least two test cases. This implies that participants did not always observe enough differences to rank the tests in terms of readability. Overall, we observe a variety of scores and evaluation patterns, e.g., participants 15, 16 \new{and 39} used almost the full scale of values while  5, 12 and 22 focused on a few scores. This seems to confirm that readability assessment is a highly subjective process, and therefore, difficult to automate without human intervention.

\begin{figure}[ht]
\centering
\includegraphics[width=0.49\textwidth]{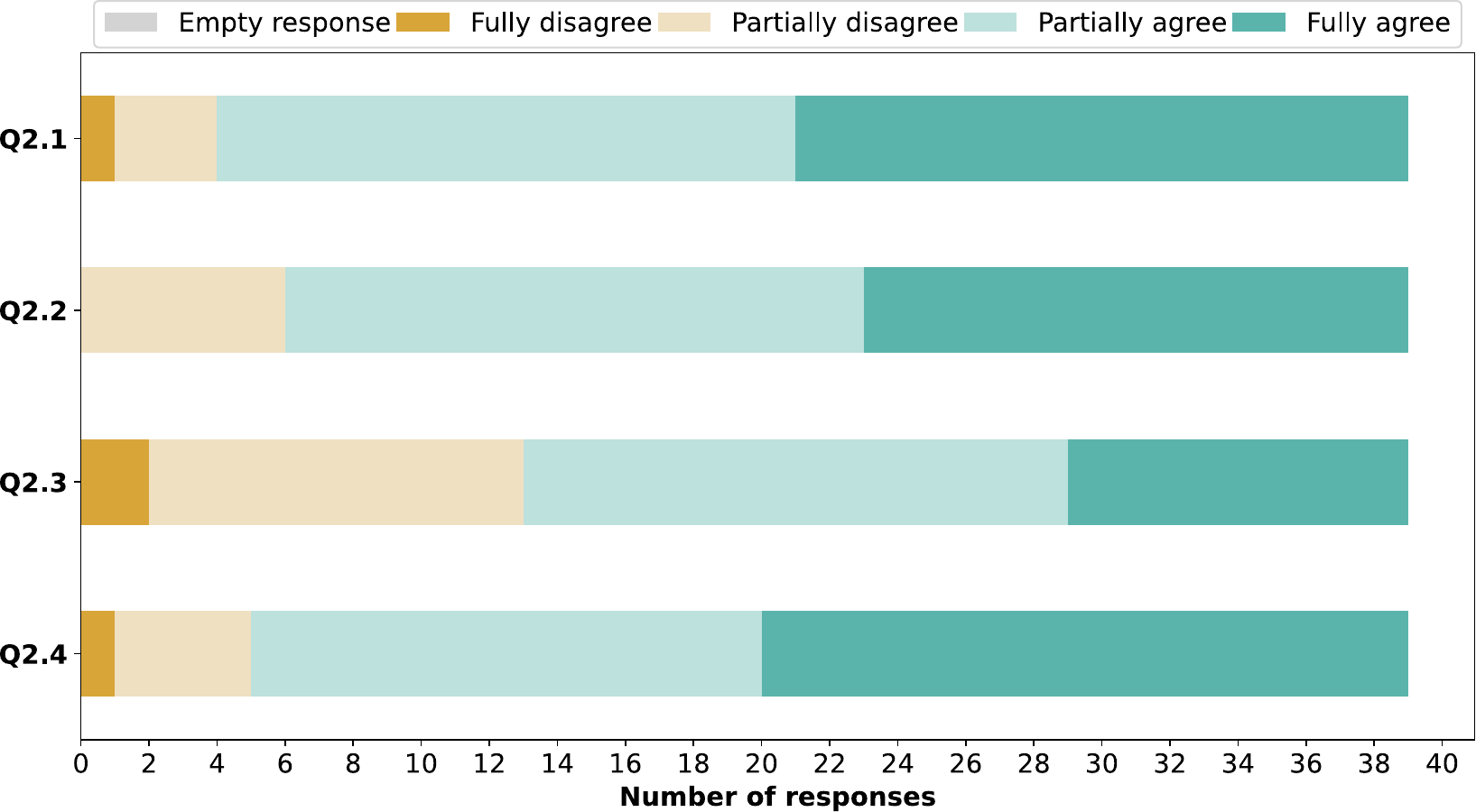}
\caption{Responses to questions related to the test readability.}
\label{fig:responses-readability}
\end{figure}

Several questions in the survey allow us to complement the analysis of RQ2, and Figure~\ref{fig:responses-readability} shows the responses to them. Although the target class is not too complex and most of the participants (\new{89}\%) said that the tests were easy to understand (Q2.1), some of them (\new{33}\%) expressed that assigning a readability score was not an easy task (Q2.3). Therefore, it could be helpful to assess the tests based on some predefined criteria, providing some guidelines for the scoring process. Focusing on the information provided during interactions, \new{85}\% of the participants perceived differences in terms of test readability (Q2.2), while \new{87}\% of them agreed that knowing the target being pursued was useful. These responses give us confidence about the suitability of using the target as the driver of the selection process, as it allows making a connection between the test cases shown.

\begin{figure}[ht]
\centering
\includegraphics[width=0.4\textwidth]{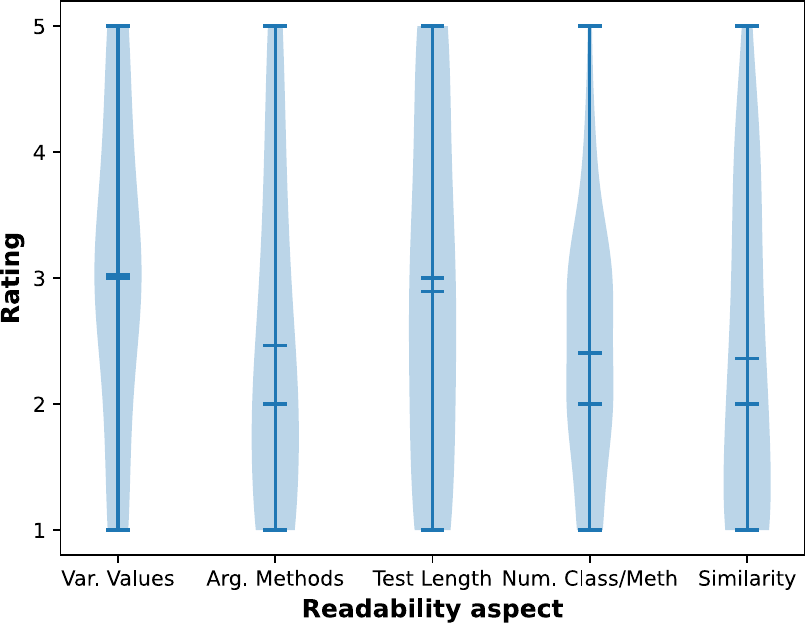}
\caption{Ratings (1=Most important, 5=Less important) assigned to each readability aspect.}
\label{fig:responses-readability-aspects}
\end{figure}

Considering the variability of scores, each participant could have thought of different code features as relevant when assessing the tests. To shed light on this, we asked them to rate whether the following aspects were important in their decision-making (some of them inspired by previous studies~\cite{Daka2015}): 1) \emph{values given to variables}; 2) \emph{arguments used in the invocation to the constructors and methods}; 3) \emph{test length (lines, characters, etc.)}; 4) \emph{number of different classes and methods appearing in the test case}; and 5) \emph{similarity to the test case he/she would have written}. Figure~\ref{fig:responses-readability-aspects} shows their responses for each readability aspect, where the horizontal lines correspond (from bottom to top) to the minimum, median, average, and maximum value. The shadow area represents the distribution of the ratings. Similar to what happened with the scores, all readability aspects were deemed as the most relevant by some participants, but the less relevant to others. Test length was not as important as could have been thought; only \new{8} participants (\new{20.5}\%) ---3 of them with more than 7 years of testing experience--- assigned it the highest rating. However, the selected tests were not excessively large in any case, so this might explain why this aspect had little influence. Arguments used in invocations and similarity to the test they would have written had more impact on their mind. Participants were also allowed to suggest new aspects and rate them. Analyzing their responses, the direct invocation to the method under test was relevant for \new{4} participants. Also, the presence of comments ---especially if the test raises exceptions--- was highlighted by 2 participants, while others were influenced by some other factors, such as the complexity of the test case, how narrow it was in the functionality being checked, and the presence of duplicated code. All these additional aspects received high ratings (1-2). 

\begin{center}
\begin{tcolorbox}
  [colback=white,colframe=gray!75!black,fonttitle=\bfseries,width=9cm]
\textbf{RQ2 answer:} 
\emph{The scores and feedback provided by the participants suggest that readability assessment is a highly subjective process based on different readability criteria. Similarity to human-written code and meaningful arguments were deemed as more relevant than test length.}
\end{tcolorbox}
\end{center}

\subsection{RQ3: Interaction process}
\label{subseq:rq3}

Table~\ref{tab:readability-results} shows a summary of the number of test cases (in total and per interaction) and targets/methods selected by \acronym in each participant's execution. The tool stopped the maximum number of interactions allowed (10) in all executions, asking participants to inspect between 15 and 28 test cases, with a median of 2 per interaction. The total number of explored targets varied between 3 (only 1 participant) and 10 (up to 4 participants), with a median of \new{8}. This suggests that \acronym requires breaking ties for a variety of targets and most of the participants were able to see test cases associated with different targets. As explained in Section~\ref{subsec:interaction-moments} and being $Max\_targets\_interaction\_moment=3$, the target selection procedure had to find targets for 3 distinct methods in each interaction moment. As such, this strategy could have had an influence on the number of different methods addressed: a median of 5 and a maximum of 6. Still, participants frequently evaluated tests covering the same methods, but the target was possibly a different one each time.

\begin{table}[ht!]
\renewcommand{\arraystretch}{1.1}
\caption{Number of tests, targets, and methods appearing during interactions.}
\label{tab:readability-results}
\vspace{-0.3cm}
\centering
\scalebox{0.9}{
\begin{tabular}{|l|rr|rrr|}
\hline
& \textbf{Min.} & \textbf{Max.} & \textbf{Median}  & \textbf{Avg.} & \textbf{Std. Dev.} \\ \hline
Test cases (interaction)    & 1 & 4 & 2.0 & \new{1.95} & \new{0.90} \\ 
Test cases (total)          & 15 & 28 & 19.0 & \new{19.46} & \new{2.93} \\ 
Targets (total)             & 3 & 10 & \new{8.0} & \new{8.13} & \new{1.32} \\ 
Methods (total)             & 3 & 6 & 5.0 & \new{4.67} & \new{0.76} \\ 
\hline
\end{tabular}
}
\end{table}

\begin{figure}[ht]
\centering
\includegraphics[width=0.4\textwidth]{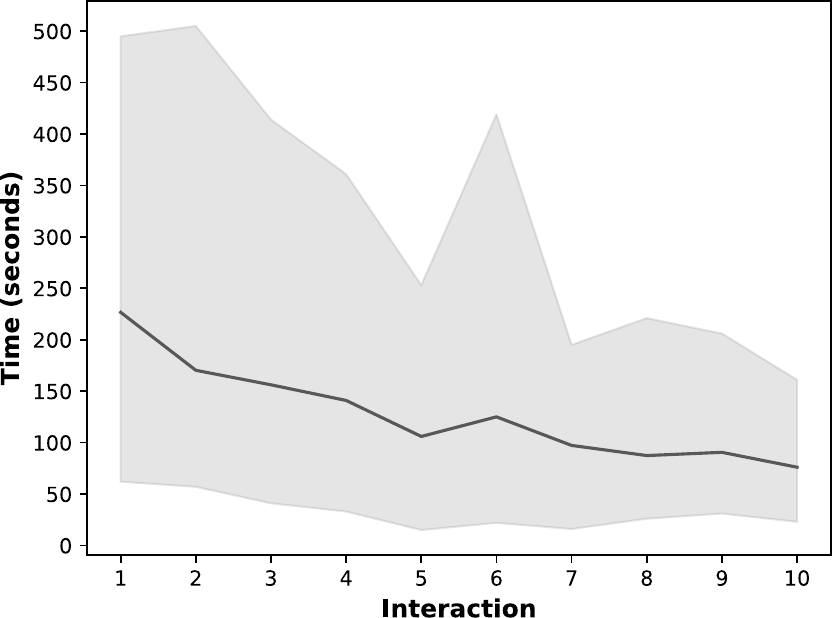}
\caption{Time spent on each interaction. The solid line represents the average among all participants. The shaded area delimits the minimum and maximum values.}
\label{fig:interaction-time}
\end{figure}

As for the interaction time, participants took between 9 and \new{37} minutes (an average of \new{21.5} minutes) in the revision of both test cases and complementary files (target and preference archive). Given that the experiment demands participants to repeat the same task in each interaction, it is interesting to study whether the time spent on each interaction remained similar or not. Figure~\ref{fig:interaction-time} shows the time in seconds dedicated (on average) by participants per interaction. The time clearly decreases as the execution progresses, which is a common phenomenon in iSBSE experiments~\cite{Ramirez2018}. The main causes are the inherent fatigue and the repetitive nature of the process, leading humans to pay more attention at the beginning. As evidence of this, \new{20} participants spent more time in the first interaction than in any other. The number increases to \new{33} participants if we take into account the 3 first interactions. Also, it seems that the task became more predictable after the first interaction, which is reflected in a significant time reduction between the first and second interaction in comparison with other moments of the execution. Another factor influencing the time curve is the number of test cases under evaluation, which depends on the minimization process and the existence of already-valued tests in the readability archive. Overall, one might expect to see a decrease in the number of unseen test cases as this archive grows. Indeed, our log data indicates that, in the first interaction, all participants evaluated 2 or 3 test cases. In the middle of the process (interaction 5), \new{15} participants were asked to assess only one new test case. In the last interaction, such a situation happened to \new{22} of them. Even so, we also found some executions differing from this behavior, in which participants still saw 3 or 4 different tests in the last interactions. This mainly happened when the algorithm was not able to find new candidates for targets/methods already explored and, as a result, focused on other targets not addressed yet.

\begin{figure}[ht]
\centering
\includegraphics[width=0.49\textwidth]{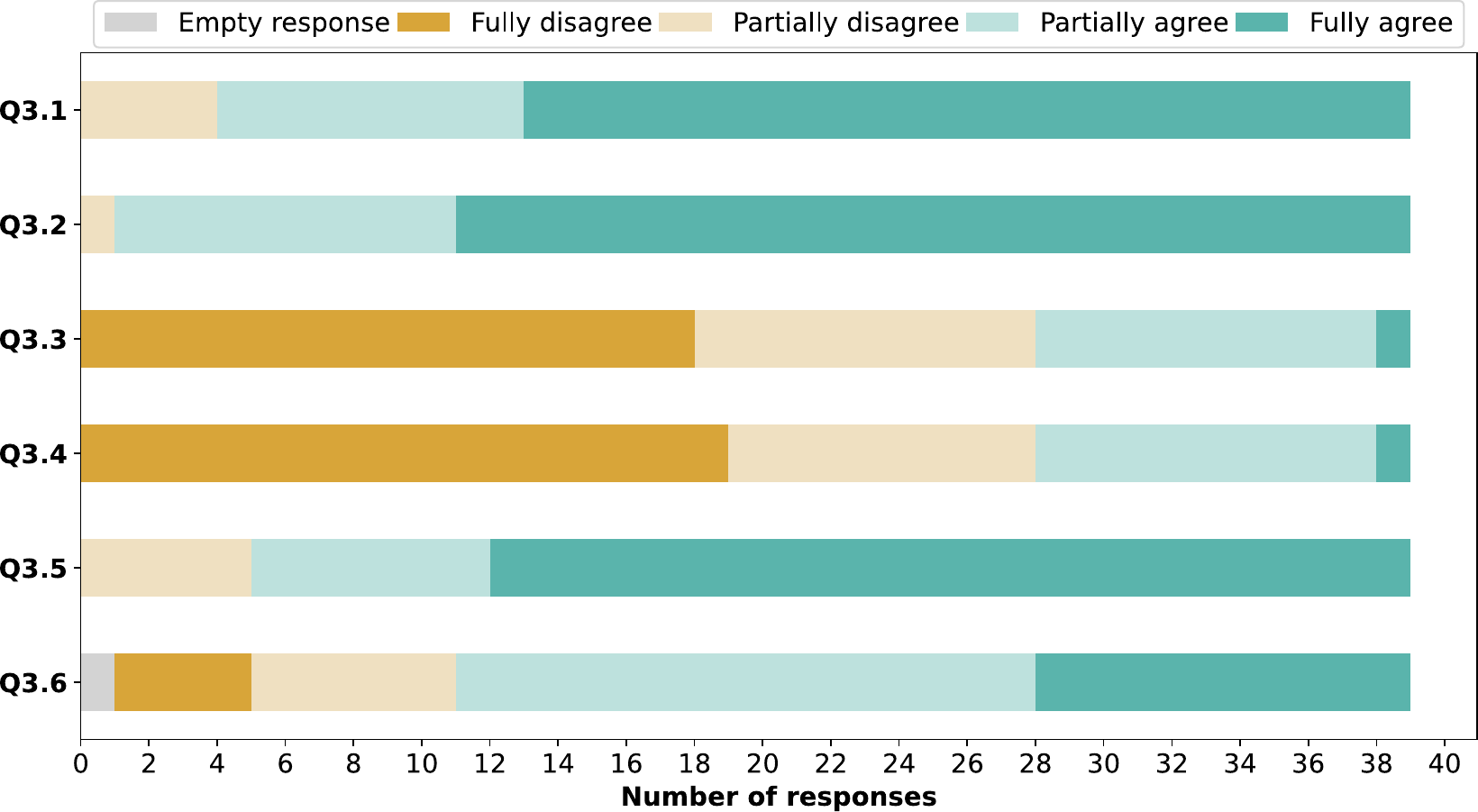}
\caption{Responses to questions related to the interaction process.}
\label{fig:responses-interaction}
\end{figure}

We contrast the information extracted from log files with the participants' responses to the questions related to the interaction process (see Figure~\ref{fig:responses-interaction}). According to their answers, the effort required in terms of the number of interactions and tests to be revised was appropriate (Q3.1 and Q3.2). We observe some discrepancy regarding the time spent on the interaction (Q3.3), with \new{28}\% of participants acknowledging that they would have preferred a shorter process. This fact could be influenced by the specific choice of targets and test cases done by \acronym, since some participants mentioned that they saw very similar tests and some repeated targets. Question Q3.4 provides interesting insights when compared with Figure~\ref{fig:interaction-time}: \new{72}\% of the participants ---partially or fully--- disagreed with the idea that they paid more attention to the earlier interactions. Despite the many factors influencing the interest and the time spent (learning effect, number of test cases, etc), it is reasonable to believe that earlier interactions demand more effort and imply more interest, and this is somehow reflected in more time spent at the beginning by all participants (almost 4 minutes on average). \new{Fatigue due to the repetitive assessment process could be another factor, but it cannot be empirically assessed from the collected data.} To provide an example of the complex relation between interest, time and effort, we can mention the case of one participant who required around 8 minutes to evaluate 2 test cases in the first interaction. Then, this participant evaluated 3 and 4 tests in around one minute during interactions 8 and 9, and finally spent more than 1.5 minutes  inspecting only one test in the last interaction (for a new target but related to a method already considered several times before). Finally, knowing the scores previously assigned (Q3.5) was seen as interesting by \new{87}\% of the participants. The preference archive, however, had a more limited utility (Q3.6). Probably, participants were more aware of the previous scores because they are directly printed on the console. The archive, in contrast, is dumped into a file, requiring users to open it. In fact, one participant (represented as ``empty response'') mentioned that he forgot the possibility of checking the archive during the process.

\begin{center}
\begin{tcolorbox}
  [colback=white,colframe=gray!75!black,fonttitle=\bfseries,width=9cm]
\textbf{RQ3 answer:} \emph{Overall, the way \acronym prepares and presents the information for the interactions seems suitable for the purpose of readability assessment. The interaction time varied among interactions and participants, decreasing as the execution progressed and influenced not only by the number of tests to revise but also by the appearance (or not) of new targets/methods.}
\end{tcolorbox}
\end{center}

\subsection{RQ4: Usefulness}
\label{subseq:rq4}

\begin{figure}[ht]
\centering
\includegraphics[width=0.49\textwidth]{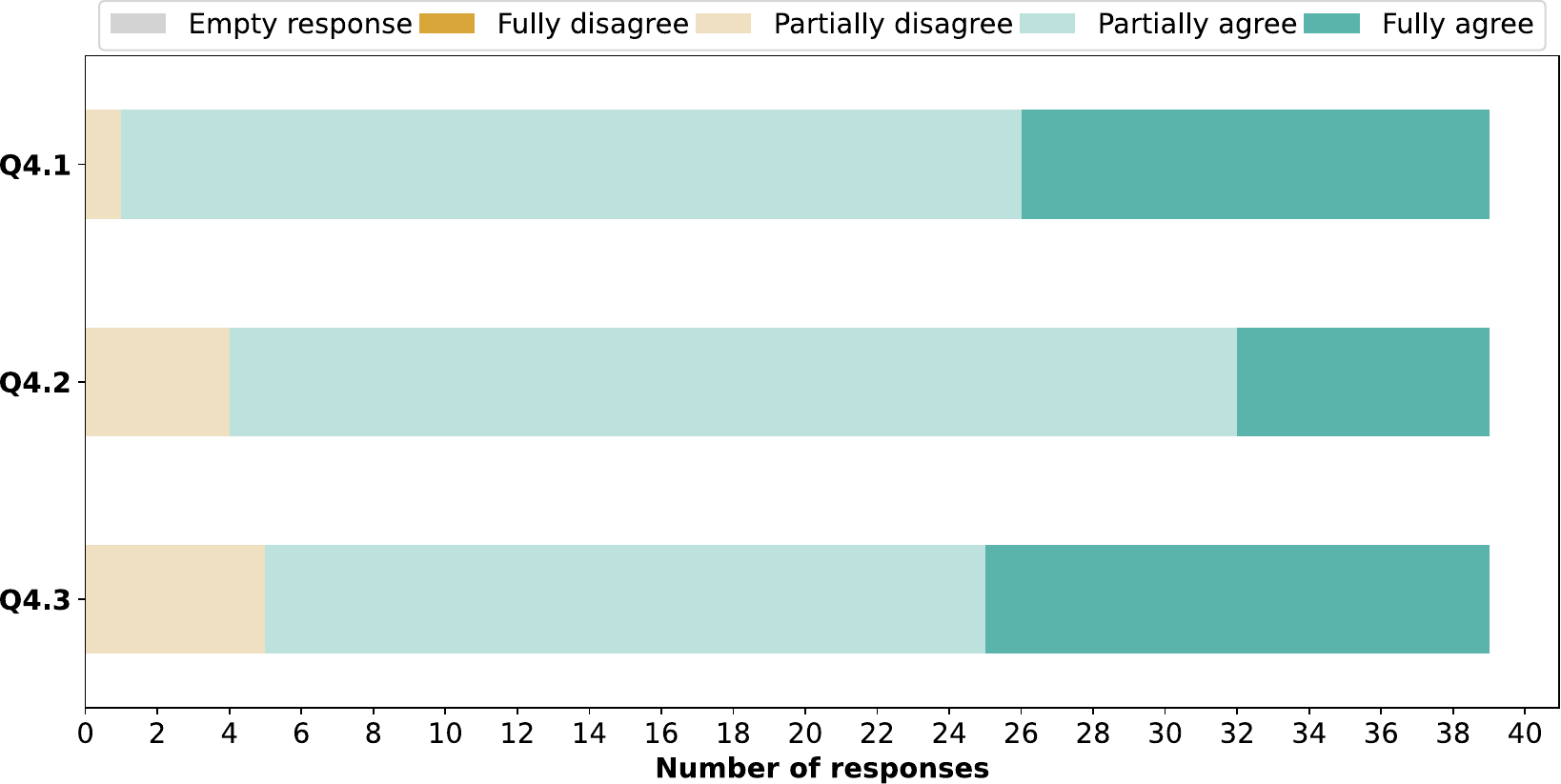}
\caption{Responses to questions related to the approach usefulness.}
\label{fig:responses-usefulness}
\end{figure}

To respond to RQ4, we analyze the responses to the last three questions (see Figure~\ref{fig:responses-usefulness}). Positive answers (partially or fully agree) clearly predominate, and none of the participants expressed strong disagreement with any of the statements. From the answers to Q4.1, we believe that the active interaction with an SBST tool helps to understand the automatically-generated tests. The possibility of inspecting intermediate candidates and the targets behind their generation may lead users to early see what types of test cases the tool handles internally and what to expect from the process.

As for Q4.2, most of the participants (\new{90}\%) partially or fully agreed that the resulting test suite was more readable than the two automatically-generated test suites. Interactions may help to mentally map the test cases found in the final test suite with their underlying purpose. Indeed, one participant suggested the inclusion of a comment in each test case indicating the covered targets for the sake of traceability. To create the final test suite, \acronym prioritizes tests with high readability scores, meaning that the test suite reflects the participant's preferences. Such a prioritization usually leads to larger test suites compared to those automatically generated, such as the test suites used for comparison. This happened for \new{34} out of \new{39} executions with respect to the number of test cases in the first test suite, and \new{24} executions for the second test suite. It is interesting that most of the participants found the resulting test suite more readable despite this increase in size. However, this test suite also contains some other unseen test cases since users only interact with a subset of the targets. These ``new'' tests presumably look similar to those in the automatically-generated test suites and might explain why most of them only perceived partial differences in terms of readability.

The majority of them (\new{87}\%) were satisfied with the overall experience (Q4.3). We can highlight that all professional developers agreed that the time spent compensated for the result\new{, with 42\% of them fully agreeing}. None of the 5 participants partially disagreeing with the benefit obtained for the time invested on the task reported any specific reason under free comments. Only one of them indicated that he missed the direct invocation of some of the methods being tested. Cross-checking their responses to other questions, 4 of them said it was not easy to assign scores (Q2.3), and 2 of them acknowledged that the process was costly in time (Q3.3) and their interest decreased over time (Q3.4). Keeping engagement with the process might be related to the variety of tests selected for assessment. The current selection strategy is based on the targets explored so far, and such a number differs from one execution to another, as discussed in Section~\ref{subseq:rq2}. New mechanisms to select targets from a greater number of methods and to widen the diversity of the presented test cases should be analyzed in future studies.

\begin{figure}[ht]
\centering
\includegraphics[width=0.4\textwidth]{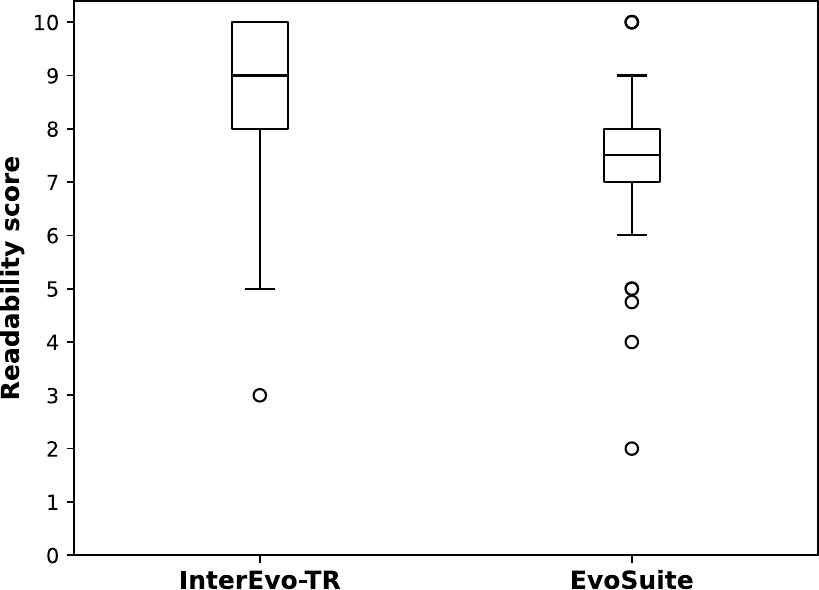}
\caption{\new{Distribution of scores for tests created by \acronym/EvoSuite.}}
\label{fig:comparison-boxplot}
\end{figure}

\new{Finally, the posterior evaluation of test cases selected by \acronym and EvoSuite shows that participants actually perceive differences in the readability of the tests chosen by these tools. Regarding the comparison (1) \emph{interactive vs interactive}, two thirds of the participants followed a similar evaluation pattern when assessing the pair of tests assigned to them, that is, they favored the same test case, exactly as the original participant did during the interaction. In contrast, 19\% of them had the contrary perception and the other 13\% gave the same score to both tests (a tie). As for (2) \emph{interactive vs automatic}, 74\% of the participants preferred the interactive test case (highly ranked by another participant) to the similar test automatically selected by EvoSuite. Only 10\% of them showed preference for the automated test case. Remarkably, in both previous comparisons the results in favor of the best-rated tests were quite spread across the 10 pairs of test cases generated for the study, revealing that the concrete selection of pairs did not have a decisive influence. Figure~\ref{fig:comparison-boxplot} shows the distribution of the readability scores for each tool. 
The comparison (3) \emph{automatic vs automatic} was successful in serving as a control mechanism. For 87\% of the participants, the difference in score for both automated tests was 2 or fewer points; those similar ratings show that they were not biased by the intuitive belief that the test cases in the pair could come from different sources, or that one of the presented test cases should be better than the other. In fact, we observe a broader range of differences in score in the comparisons (1) and (2). As an example, 48\% of the participants gave differences in score of 3 points or more when confronting the pair of tests \emph{interactive vs interactive}. } 

\begin{center}
\begin{tcolorbox}
  [colback=white,colframe=gray!75!black,fonttitle=\bfseries,width=9cm]
\textbf{RQ4 answer:} \emph{Most of the participants were very positive regarding the interactive experience and approach usefulness, perceiving differences with respect to fully automated results in terms of readability. All professional developers agreed that the result obtained was worth the time invested.} 
\end{tcolorbox}
\end{center}

%-----------------
% 6 - Discussion
%-----------------

\section{Discussion}
\label{sec:discussion}

In this section, we discuss additional aspects related to the interactive experience and the behavior of \acronym, based on the participants' responses to open questions and the impact of design decisions of the approach, respectively. 

\vspace{2pt}
\noindent\textbf{\emph{Actions for readability improvement and comments.}} Participants gave their opinion about other interactive opportunities they would have liked to have had in addition to providing readability scores. Among the five proposed possibilities, rewarding/penalizing individual parts of the test cases (\new{56}\%) and modifying values of variables and arguments (\new{54}\%) were the two options most consistently selected by participants. Combining parts of two or more test cases (\new{38}\%), suggesting the inclusion of new objects or the invocation of different methods (\new{28}\%) and adding new assertions (28\%) were also marked but to a lesser extent.

The different free comments were numerous, including suggestions and issues they faced. Among them, some participants mentioned that, in some interactions, the shown test cases were too similar (e.g., only the value of an argument changed) or not complex enough. This might also explain why test length was not perceived as highly relevant. A possible solution to this would imply incorporating a mechanism to show both diverse and not too simple test cases, e.g., based on some objective measures. Other participants indicated that some test presentations did not feel natural, e.g., negative values in parentheses or too high values passed as arguments. Sometimes they would recommend splitting a test case into two. These are issues mainly related to the internal operation of the test generation process. Two of them described some problems when providing the scores (they failed when introducing the values and could not undo the action), suggesting that they would like to re-evaluate already seen tests. \new{Based on this, we decided to implement the option of revisiting candidates in the tool (\emph{revisit\_candidates}). With this option enabled, the tester can provide a new score to tests already shown in previous interactions when they appear again.} Developing an intuitive GUI could help solve these issues as well.

\vspace{2pt}
\noindent\textbf{\emph{Preparation time.}} Preparing interactions involves the selection of candidate tests, their minimization and the addition of assertions, which adds an overhead to the search. We should note that the minimization process is only applied when at least two candidates covering the target are found; otherwise, the next target in the list is selected and the process is repeated until the number of interactions for that interaction moment is finally reached (as described in Section~\ref{subsec:incorporation}). Moreover, the selected tests are only augmented with assertions once it is guaranteed that the interaction will take place, i.e., when some minimizations are found whose readability evaluation is still pending. As a result of this procedure, the mean preparation time in the experiment was \new{11.7} seconds (around 1 second per interaction), which we judge acceptable when compared to the search (2 minutes in our case) and the interactions themselves (\new{21.5} minutes on average). Whether this preparation time significantly increases with other classes should be further investigated. 

\vspace{2pt}
\noindent\textbf{\emph{Minimization and readability archive.}} We should recall that, after the minimization of inner test cases, we can find redundant minimized tests, thereby reducing the number of candidates selected initially. Thus, considering all attempts to produce an interaction with at least two candidates, the minimization process halved the size of the set of selected test cases, i.e., \new{47.9}\% turned out to have the same minimization as another candidate. Furthermore, this reduction spoiled \new{42.5}\% of all attempts, when only a single minimization was found and there was no point in asking for revision.  Again, this suggests that it could be helpful to increase the diversity of the candidates for the purpose of readability assessment. Then, the set of remaining minimizations was further reduced, as \new{80.6}\% of them were found in the readability archive. This canceled \new{45.1}\% attempts of interaction that only contained already seen minimizations. Therefore, only 12.4\% of the attempts succeeded in completing the interaction. In summary, the minimization process and the readability archive are effective and necessary mechanisms to avoid showing redundant tests and useless interactions, as they seem to be frequent. This, however, may depend on the complexity of the class and its particular code features.

\vspace{2pt}
\noindent\textbf{\emph{Preference archive.}} In the experiments carried out, we cannot know whether the parameter \emph{P\_Preference\_Selection} contributed to the generation of more readable test cases. 
Due to the subjective nature of this problem, we cannot assess to which extent the code features that made those test cases the most readable for the tester actually propagated to new test cases in next generations. We should remark, however, that the number of different test cases contained in the coverage archive will be usually much larger than that in the preference archive. Consequently, selecting too many tests from this archive could hamper the population diversity, so this parameter should not be set to a high probability.

%-----------------
% 7 - THREATS
%-----------------
\section{Threats to validity}
\label{sec:threats}

Experiments with participants pose some threats to internal validity due to the sample size and its representativeness. Firstly, the number of participants (\new{39}) is aligned with previous studies in SBST~\cite{Panichella2016, Roy2020}, and clearly superior to what is frequent in iSBSE~\cite{Ramirez2019}. In this sense, we consider that the sample size is adequate for the type of qualitative analysis presented. Notice that increasing the sample size would not necessarily imply more agreement given the subjective scoring process. Despite this, most of the participants were not experts in software testing, so the results might not be representative of what professional testers would think or do. However, we were able to recruit several professional developers, and the type of testing task (unit testing over a Java class) does not really require such a kind of testing expertise. To overcome the possible lack of knowledge in relation to the task, we devoted some time to explaining basic concepts and running an example. Even though we cannot be completely sure whether they understood the task and paid enough attention to its resolution, their positive responses give us enough confidence about their commitment. 

In the study, \acronym could not be compared with other previous interactive approaches ---as far as we know, there is no comparable proposal of interactivity for readability assessment. In this sense, the perception of usefulness might stem from the good performance of EvoSuite in generating automated tests more than from the interactive approach, since most of the participants had not been in touch with SBST tools in the past. Even if this is the case, the fact that they mostly found the interactive process appealing and useful offers positive evidence \emph{per se}, especially considering the burden of manual code inspection. \new{As in similar experiments with participants, the social-desirability bias also poses a threat to the validity. For instance, it is possible that most of the readability scores are medium to high because participants thought we had added ``some more readable tests"  in the tests shown for revision. However, participants did not exactly know what were the  aspects under evaluation, as they were only told to rate the extent to which the presented test cases looked readable for them, and how that information was used by the tool. Therefore, ultimately, they may have biased the results in favor or against our interests. \new{In the analysis of usefulness, we added control questions were both test cases in the pair had been generated by the same tool (\acronym and EvoSuite) to avoid any guesses about the purpose of the comparison.}}

Our experiment was planned using a single class under test. Interactions demand considerable cognitive effort, so asking participants to perform a second run with another class could have compromised the results due to a learning effect. To alleviate this threat, all participants executed the tool using a different seed and, therefore, they could observe different targets and/or tests. To avoid information overload, test cases were generated with respect to three accessible coverage criteria (Line, Branch and Weak Mutation). Enabling some additional coverage criteria can lead to the generation of more tests to cover those additional targets. This could require a greater number of interactions to make a proportional impact on the test suite to that achieved with those three criteria. However, we have focused on three frequently used criteria~\cite{Rojas2017-Whole}. Also, we conjecture that a broader set of criteria could result in turn in more diverse targets and test cases being selected, which could ultimately make the process even more interesting for expert testers.

%-----------------
% 8 - Related work
%-----------------

\section{Related work}
\label{sec:related}

Next we summarize related works on 1) test readability in automated testing and 2) interactive approaches in SBST.

\vspace{2pt}
\noindent\textbf{\emph{Test readability.}} As noted by Rojas et al.~\cite{Rojas2017}, generating more readable test cases represents an open challenge derived from its intrinsic subjective character, which makes it difficult for test generation tools to achieve more human-competitive results~\cite{arcuri2018experience,Shamshiri2018}. In this line, Daka et al.~\cite{Daka2015} collected human-rated test cases to train a regression model and extract desirable code properties, such as identifier length or code constructs like loops or assertions. The readability estimation provided by the predictive model can be used to automatically analyze test cases in a post-processing step, or as part of a multi-objective formulation~\cite{Daka2015ssbse}. From their experiments with participants, the authors highlight that more readable test cases allow testers to make decisions faster, but some aspects (e.g., variable names) still need to be improved. Other empirical studies have also concluded that understanding the intention of automatically-generated test cases requires considerably more time, being limited by the lack of comments and descriptive names~\cite{Shamshiri2018}.

Several authors have studied how automated methods could overcome the aforementioned issues. Daka et al.~\cite{Daka2017} tackled the generation of meaningful test method names that synthesize the behavior of the code under test, showing that the automatically-generated names were as descriptive as manual ones. Panichella et al.~\cite{Panichella2016} proposed TestDescriber as an approach to improve the understandability of automated test cases by means of individual summaries. In their evaluation, these summaries helped developers both to find twice as many bugs and to better comprehend the purpose of test cases. Recently, Roy et al.~\cite{Roy2020} presented DeepTC-Enhancer, an approach based on deep learning and templates to enhance both the documentation of the tests and additional code aspects, including variable renaming. In their study, participants reported a significant improvement in readability when applying DeepTC-Enhancer with respect to previous methods~\cite{Daka2017,Panichella2016}. In contrast to these works, our approach does not aim at improving the readability of the test suite by enhancing different aspects of its appearance. Instead, our interactive approach seeks to modify the composition of the test suite itself by replacing test cases with those satisfying the tester's preferences. All previous developments differ from our approach in that they operate automatically in the end, i.e., without any human intervention. \new{Also, \acronym does not modify the test cases themselves, but the opinion of the tester takes part in the selection of test cases for the final test suite. As such, we claim that both, interactive and automated approaches, could be used in combination to boost the readability and acceptance of the results. On the one hand, the interactive execution would add the necessary human perspective in the selection of tests. On the other hand, fully automated techniques would help improve the overall appearance of the final test suite by showing a cleaner test code. These mechanisms could be invoked before presenting the test cases to the testers, helping them when executing \acronym with more complex classes. In such situations, the test cases selected for revision can really benefit from additional comments and renaming improvements.}

\vspace{2pt}
\noindent\textbf{\emph{Interactive SBST.}} Marculescu et al. analyzed the possibilities that interactive evolutionary computation could bring to SBST and, in particular, to the generation of test data in black-box testing~\cite{Marculescu:SBSE12}. Their original idea was to allow expert testers to evaluate some solutions and interact with the tool by incorporating their knowledge and understanding of the SUT. The following works by the same authors~\cite{Marculescu2013icmla,Marculescu2013} explored in detail the idea of ``user as fitness function'', where domain specialist was asked to periodically re-assign weights for the objective functions of a differential evolution algorithm. In their algorithm, applied to an embedded software from an industrial partner~\cite{Marculescu2015isbst}, candidates are encoded as numerical arrays that contain the inputs to the SUT, and up to 11 objectives are defined to evaluate its behavior. Among other aspects, they motivated the need of providing clear and accurate information during the interaction and observed differences in how domain specialists prefer to interact. The empirical study also revealed that the interaction greatly contributed to exploring new regions of the search space, though the resulting tests were not as good as manual tests. Their last work presents an updated prototype of the interactive SBST tool focused on fault detection, together with an experience report of its transfer to industry~\cite{MarculescuFTP18}. We share with these works the application of interactive optimization to an SBST problem, though the approaches are notoriously different. We focus on white-box testing for object-oriented software (as opposed to black-box testing), which implies that our interaction demands inspecting source code instead of test data or SUT behavior. Therefore, the approach is different in both the purpose of the interaction and the feedback integration, since our aim is not to adjust the fitness function to guide the search but to complement it with preferences about test readability.

%-----------------
% 9 - CONCLUSION
%-----------------

\section{Conclusion}
\label{sec:conclusion}

Search-based test generation has been extensively studied in the past, and now automatic tools like EvoSuite provide efficient evolutionary methods to achieve high coverage. However, the generated tests are still difficult to understand by testers, who cannot participate in the process. In this work, we propose \acronym, an interactive approach that lets testers evaluate the readability of candidate test cases during the search. Readability assessment at periodical steps has been integrated into DynaMOSA (EvoSuite's default algorithm), as well as several mechanisms to prepare the interactions and integrate human feedback. Our experiment with participants reveals an assortment of results regarding how they faced test readability assessment, how they behaved during interactions and how useful they perceived this process. We highlight that \acronym is able to present relevant information to testers, whose feedback is reflected in the results. These are positive signs towards a wider adoption of SBST tools and the acceptance of automated results. Also, we consider that \acronym can be helpful for students and junior developers to introduce themselves in automated testing.

In this study, the tester could provide a single readability score during interactions. As future work, we plan to analyze and implement further interactive opportunities, such as the possibility to modify the candidates in addition to assessing them. Moreover, these new opportunities could enhance the capability to detect challenging faults as testers could provide their knowledge of the internal structure of the source code. Based on the participants' feedback, \acronym could also be improved in terms of tool usability. \new{Being a human-based approach, it would be also interesting to analyze possible cognitive biases usually affecting software engineering tasks~\cite{Mohanani2020}, especially those action-oriented and related to decision-making.} \new{In this sense, InterEvo-TR could be extended to allow the joint evaluation by several testers as way to reduce inaccuracies coming from less expert testers. In addition, it would be interesting to compare our interactive approach against the manual design of test cases, thus analyzing whether interactive test generation tools could help testers save time. In the long term, we plan to study whether an interactive approach can actually promote a wider adoption and acceptance of SBST from the software industry.}

\section*{Acknowledgments}
The authors would like to thank the participants for their valuable time, strong interest and positive feedback, and the \'Area de Sistemas de Informaci\'on in the University of C\'adiz. This work was partially supported by the European Commission (FEDER), the Spanish Ministry of Science and Innovation (projects PID2021-122215NB-C33, RTI2018-093608-BC33, RED2018-102472-T, Grant PID2020-115832GB-I00 funded by MICIN/AEI/10.13039/501100011033) and the Andalusian Regional Government (postdoctoral grant DOC\_00944).

\bibliographystyle{IEEEtran}
\bibliography{references}

\begin{IEEEbiographynophoto}{Pedro Delgado-P\'erez} received the Ph.D. degree in computer science engineering (2017) and works as Assistant Lecturer in the University of  C\'adiz (Spain). He has mainly centered his research on testing techniques. His research interests also include SBSE and OO programming.
\end{IEEEbiographynophoto}

\begin{IEEEbiographynophoto}{Aurora Ram\'irez}
received the PhD in Computer Science from the University of C\'ordoba (Spain) in 2018, where she is currently a postdoctoral researcher. Her research interests include SBSE, software analytics, interactive optimization and explainable artificial intelligence.
\end{IEEEbiographynophoto}

\begin{IEEEbiographynophoto}{Kevin J. Valle-Gómez}
currently works as Assistant Lecturer at the University of Cadiz (Spain). His research interests focus on the improvement of software testing processes in line with the needs of industry.
\end{IEEEbiographynophoto}

\begin{IEEEbiographynophoto}{Inmaculada Medina-Bulo} received her PhD in computer science at the University of Seville (Spain). She has worked in the Department of Computer Science and Engineering of the University of C\'{a}diz (Spain) since 1995. She is the main researcher of the UCASE Research Group. Her research interests focus on Software Testing, SBSE, IoT and CEP. 
\end{IEEEbiographynophoto}

\begin{IEEEbiographynophoto}{Jos\'e Ra\'ul Romero} 
received his PhD at the University of M\'alaga (Spain), and is with the Dept.  of Computer Science of the University of C\'{o}rdoba. His research interests focus on democratization of data science and the application of intelligent systems to software engineering. 
\end{IEEEbiographynophoto}

\end{document}